\documentclass[
  reprint,
 superscriptaddress,
 amsmath,amssymb,
 aps,
pra,
]{revtex4-2}
\usepackage{orcidlink}
\usepackage{graphicx}
\newcommand\sbullet[1][.5]{\mathbin{\vcenter{\hbox{\scalebox{#1}{$\bullet$}}}}} 
\usepackage{physics}
\usepackage{xspace}
\newcommand{\ps}{phase space\xspace}
\usepackage{hyperref}
\usepackage{color}
\newcommand{\VEC}[1]{{\mbox{\boldmath${#1}$}}}

\usepackage[normalem]{ulem}

\makeatletter
\newsavebox{\@brx}
\newcommand{\llangle}[1][]{\savebox{\@brx}{\(\m@th{#1\langle}\)}%
  \mathopen{\copy\@brx\kern-0.5\wd\@brx\usebox{\@brx}}}
\newcommand{\rrangle}[1][]{\savebox{\@brx}{\(\m@th{#1\rangle}\)}%
  \mathclose{\copy\@brx\kern-0.5\wd\@brx\usebox{\@brx}}}
\makeatother

\usepackage{forloop,ifthen} 

\usepackage{xifthen}
\newcommand{\refAppendix}[6]{#1
  \ifthenelse{\isempty{#2}}%
    {}
    {\protect\cite{#2}}
    #3\protect\ref{#4}#5#6\xspace
}

\begin{document}

\title{Wigner's Phase Space Current for Variable Beam Splitters
  \\
  --- Seeing Beam Splitters in a New Light --- }

\author{Ole Steuernagel\orcidlink{0000-0001-6089-7022}}
\email{Ole.Steuernagel@gmail.com}
\affiliation{Institute of Photonics Technologies, National Tsing Hua University, Hsinchu 30013, Taiwan}

\author{Ray-Kuang Lee\orcidlink{0000-0002-7171-7274}}
\email{rklee@ee.nthu.edu.tw}
\affiliation{Institute of Photonics Technologies, National Tsing Hua University, Hsinchu 30013, Taiwan}
\affiliation{Department of Physics, National Tsing Hua University, Hsinchu 30013, Taiwan}
\affiliation{Center for Theory and Computation, National Tsing Hua University, Hsinchu 30013, Taiwan}
\affiliation{Center for Quantum Science Technology, Hsinchu 30013, Taiwan}
 
\date{\today}
\begin{abstract}
  Beam splitters allow us to superpose two continu\-ous single mode quantum systems. To study the
  behaviour of their strongly mode mixing dynamics we consider variable beam splitters and their
  dynamics using Wigner's \ps distribution,~$W$, the evolution of which is governed by the
  continuity-equation $ \frac{\partial}{\partial \tau} W = - \VEC{\nabla } \cdot \VEC{J}$. We derive
  the form of the corresponding Wigner current,~$\VEC{J}$, of each outgoing mode after tracing out
  the other. The influence of the modes on each other is analyzed and visualized using their
  respective Wigner distributions and Wigner currents. This allows us to perform geometrical
  analyses of the mode interactions, casting new light on beam splitter behaviour. Several of the
  presented results should be immediately testable in experiments.
\end{abstract}

\maketitle

\section{Introduction}

In quantum optics beam splitters are often used to entang\-le two matched
modes~\cite{Titulaer_Glauber__PR66} in order to study the behavi\-our of one outgoing mode
conditioned on the state of the other.  Such studies traditionally focus on the states alone,
typically in Fock-space or using \ps
representations~\cite{Campos_PRA89,Leonhardt_PRA93,Dakna__EPJD98}.

Here, we show that, instead of focusing on the state alone, it can be useful to complement such
studies by describing and visualizing the dynamics of the beam splitter interaction using Wigner's
\ps
current,~$\VEC{J}$,~\cite{Bauke_2011arXiv1101.2683B,Ole_PRL13,Oliva_Kerr_18,Braasch_PRA19,Chen_PRA23},
thus presenting beam splitter behavior in a new light.

Wigner's description of quantum systems~\cite{Wigner_PR32} in phase space, based on Wigner's
distribution $W(x,p)$, has given us visualizations comparing classical with quantum states~\cite{Berry_JPA79,Korsch_PD81,Leonhardt_PQE95,Schleich_01,Zurek_NAT01,Zachos_book_21}.  Moreover, the
associated Wigner current, $\VEC{J}$, can be constructed from a variable beam splitter's fictitious
time evolution~\cite{Lvovsky_16squeezed,Chen_PRA23} and allows for a direct visualization of the
system dynamics and its comparison with classical Hamiltonian
flows~\cite{Ole_PRL13,Kakofengitis_EPJP17,Oliva_Shear_19}. No such current exists to describe
$\rho$'s evolution (no one studies the commutator~$[\hat H, \hat \rho]$ by itself).

Since the state space for two optical modes is four-dimensional we trace out either mode and study
the behavior of the remaining mode through is two-dimensional Wigner-distribution and -current.
This allows us to study and visualize the behaviour of such systems using the \ps current vector
fields, $\VEC J$, and their `quantum phase portraits' (as two-dimensional plots).  Tracing out one
mode, we derive the expressions for the `effective Moyal
brackets'~\cite{Hancock_EJP04,Moyal_MPCPS49,Groenewold_Phys46,Zachos_book_21} and the form of
Wigner's \ps current for the other mode.

We highlight two aspects whose investigation should become experimentally
accessible~\cite{Chen_PRA23}: the inversion of the direction of the Wigner
current~\cite{Ole_PRL13,Oliva_Kerr_18} associated with the presence of \mbox{areas} in \ps where the
Wigner distribution is negative, and the pronounced violation of \ps volume
conservation~\cite{Oliva_PhysA17} in quantum dynamics.

We emphasize that superposing perfectly matched~\cite{Titulaer_Glauber__PR66} modes at a mixing beam
splitter is governed by one of the simplest coupling dynamics one can think of. By themselves,
matched modes evolve as equal harmonic oscilla\-tors and their dynamics can be factored out, whilst
their two-mode inter\-action hamiltonian is simple, bilinear in the modes, see Eq.~(\ref{eq:H_M})
below. This permits us to study how superposed beam splitter modes influence each other without the
complicating effects of a complicated system's intrinsic
dynamics~\cite{Oliva_Kerr_18,Oliva_Shear_19,Ole_JPAMT23}, allowing us to pinpoint the details of the
mechanism by which mode entanglement is responsible for non-classical
behaviour~\cite{Kim__PRA02,Wang__PRA02} observed in beam splitter interactions.

Classical mechanics has benefited enormously from the study of classical \ps
portraits~\cite{Nolte_PT10}.  In the quantum case, \ps-based approaches allow for the study of
`quantum phase portraits' (collections of momentary snapshots of field lines arising from the
integration of the vector field $\VEC{J}$~\cite{Ole_PRL13,Kakofengitis_EPJP17}). There is no other
formulation of quantum theory~\cite{formulationsQuantumTheoryfootnote} that allows for such an
intuitive way of studying quantum dynamics and is so reminiscent of classical dynamics
studies~\cite{Hancock_EJP04}.  Plotting~$\VEC{J}$ and its quantum \ps portraits allows us to apply
geometrical reasoning that sheds new light on the behaviour of beam splitters.

Sect.~\ref{sec:WignerGeneral} introduces Wigner's continuity equation and \ps current,~$\VEC{J}$, in
Sect.~\ref{sec:WignerJ2mode} we derive its form for variable beam
splitters. Sect.~\ref{sec:Examples} gives several exam\-ples and highlights interesting features of
their \ps dynamics. Subsection~\ref{sec:VolumeChanges} highlights the occurrence of singular
\ps volume changes and subsection~\ref{sec:Entanglement_J_W} shows that this is due to mode
entanglement, our Conclusions,~\ref{sec:Conclusion}, put our findings into context.

\section{Wigner distribution and its Continuity Equation\label{sec:WignerGeneral}}

At first glance it might appear that the study of conditional dynamics, since it involves sudden
`collapses' due to measurements, seems to leave no room for a continuous description by a continuity
equation with a current~$\VEC{J}$. After all, we know that Schr{\"o}dinger's equation does not
describe measurement collapses.

And yet, in experiments we often change a system parameter continuously such that instead of
physi\-cal time we study the dependence on this parameter playing the role of an `effective time' in
a fictitious evolution~\cite{Lvovsky_16squeezed,Chen_PRA23}. And we execu\-te the measurement,
inducing a collapse, after the parameter-mediated interaction is over.  In this case we can still
characterize the behaviour of a system as continuously evolving (and then being terminated by a
measurement~\cite{Chen_PRA23}).

The time-evolution of Wigner's quantum \ps distribution
$W(x,p,\tau)$~\cite{Case_AJP08,Hillery_PR84}, for a one-dimensional continuous system, is governed
by its \ps current~$\VEC{J}$ and obeys the continuity
equation~\cite{Oliva_PhysA17,Donoso_PRL01,Skodje_PRA89}
\begin{eqnarray}\label{eq:continuity} \frac{\partial W(x,p,\tau)}{\partial \tau} + \VEC{\nabla}
\cdot \VEC{J}(x,p,\tau) = 0 \; .
\end{eqnarray}
Here, $\VEC{\nabla} = (\frac{\partial}{\partial_x}, \frac{\partial}{\partial_p})$ is the \ps
divergence operator with respect to position~$x$ and momentum~$p$, $\tau$ is time,
and~$\VEC{J}=(J_x,J_p)^{\bf \intercal}$ has two components and is a functional of $W$ and the system
hamiltonian~$H(x,p,\tau)$.

We specifically study the use of beam splitters as mode mixers. Small varia\-tions in reflectivity
$r$ allow us to invoke a continuous description using an effective hamiltonian and thus continuity
equation~(\ref{eq:continuity})~\footnote{Since only the divergence of the current enters the time
  evolution equation~(\ref{eq:continuity})~\cite{Ole_23photonAddition}, there can exist ambiguity
  since purely rotational vector fields can be added to the current~$\VEC J$ and other physical
  arguments might have to be employed in order to remove this ambiguity, for an example
  see~\cite{Oliva_Kerr_18}.  Here, this problem does not occur.}.

In our approach the study of details and changes of the Wigner distributions is underpinned by plots
of their \ps current~$\VEC J$ together with~$\VEC J$'s field lines. This complements the standard
treatment of beam splitter behaviour in terms of photon statistics~\cite{Campos_PRA89} or their
effects on wavefunctions~\cite{Leonhardt_PRA93}.

\section{Current ${\VEC J}$ from Moyal's bracket~\label{sec:WignerJ2mode}}

We will now remind the reader of how Wigner's and Schr{\"o}dinger's formulation of quantum theory
are connected mathematically~\cite{Zachos_book_21}.

Consider a single-mode operator given in coordinate
representation~$\langle x-y| \hat O | x+y \rangle = O(x-y,x+y)$. To map to Wigner's \ps formulation
we employ the
Wigner-transform,~${\cal W}[\hat O]$,~\cite{Hancock_EJP04,Cohen_LectureNotes18,Zachos_book_21}
\begin{align}\label{eq:WignerWeyl_Trafo}
  {\cal W}[\hat O](x,p) =
  \int_{-\infty}^\infty dy\; O(x-\frac{y}{2},x+\frac{y}{2})\; {\rm e}^{\frac{{\rm i}}{\hbar} p y}\, .
\end{align}

If $\hat O$ is a (normalized) single-mode density matrix~$\hat \rho$, then the associated normalized
distribution in the Wigner formulation is the Wigner distribution
\begin{align}\label{eq:WignerDistribution}
  W(x,p,\tau) \equiv  \frac{ {\cal W}[\hat \rho] }{2 \pi \hbar} \; .
\end{align}

The Wigner transform of the von Neumann time evolution equation
${\cal W}\left[ \frac{\partial \hat \rho}{\partial \tau} = \frac{1}{{\rm i}\hbar} [\hat H, \hat
  \rho] \right]$ is
\begin{equation}\label{eq:moyal_motion}
  \frac{\partial W}{\partial \tau} = \{\!\!\{ {H} , W \}\!\!\} \; ,
\end{equation}
in which the Groenewold-Moyal
bracket~\cite{Groenewold_Phys46,Moyal_MPCPS49,Zachos_book_21} 
is the quantum version of the Poisson bracket with the explicit form
\begin{align}\label{EqMoyalBraket}
\!\!  \{\!\!\{ f, g\}\!\!\} 
     = \frac{2}{\hbar} f(x,p) \sin\!\!\left[\! \frac{\hbar}{2} \!\!\left( 
        \overleftarrow{\frac{\partial}{\partial x}} \overrightarrow{\frac{\partial}{\partial p}}
         - \overleftarrow{\frac{\partial}{\partial p}} \overrightarrow{\frac{\partial}{\partial x}}
    \right)\!\!\right] g(x,p) \; ;
\end{align}
where arrows indicate the `direction' of differentiation:
$f\overrightarrow{\frac{\partial}{\partial x}} g = g\overleftarrow{\frac{\partial}{\partial x}} f =
f \frac{\partial}{\partial x} g$.

Eq.~(\ref{eq:moyal_motion}) can be rewritten as the divergence of Wigner's \ps current~\cite{Oliva_Kerr_18,Oliva_PhysA17,Ole_23photonAddition},
yielding Eq.~(\ref{eq:continuity}).

\subsection{Contrast with Schr{\"o}dinger Equation\label{subsec:contrastSchroedingerWigner}}

We note that Eq.~(\ref{eq:continuity}) is the Fourier-transform~(\ref{eq:WignerWeyl_Trafo}) of the
von-Neumann equation~(\ref{eq:moyal_motion}), just as $W(x,p)$ is the Fourier-transform of the
density matrix~$\rho(x,x')$.

The Fourier-transform~(\ref{eq:WignerWeyl_Trafo}) is invertible, in fact it is a unitary map:
Wigner's and the Schr\"odinger--von~Neumann formulation are therefore \emph{unitarily equivalent}.

Additionally to the Schr\"odinger--von~Neumann formulation, Wigner's formulation of quantum theory
has the great advantage of allowing us to visua\-lize quantum dynamics in \ps more directly since
$W$ is real-valued~\cite{Hillery_PR84,Kurtsiefer_NAT97,Grangier_SCI11}. Compared to other quantum
\ps distributions~\cite{Cahill_PR69b}, Wigner's~\cite{Wigner_PR32} is
special~\cite{Royer_PRA77,Manfredi__PRE00,Wlodarz__IJTP03} and
intuitive~\cite{Grangier_SCI11,WignerSpecialfootnote}, here, we use it exclusively.

Additionally, unlike formulations of quantum physics that are not based on
\ps~\cite{formulationsQuantumTheoryfootnote}, Wigner's formulation allows us to use the continuity
equation~(\ref{eq:continuity}), and hence Wigner's \ps current~$\VEC{J}$ and its quantum phase
portraits.

Recent experimental work shows that the reconstruction of Wigner's \ps current~$\VEC{J}$ and its
effects on the evolution of the system can be studied with high resolution~\cite{Chen_PRA23}.

\subsection{Beam Splitter and its effective Hamiltonian}

We consider an ensemble of measurements performed on a system of two modes~$\hat a$ and~$\hat b$
which arise from perfectly matched incoming modes~$a_{\rm in}$ and $b_{\rm in}$ after being mixed through
interaction at a `perfect' lossless two-mode beam splitter with variable transmissivity.

{The associated unitary mode mixing operator~\cite{Campos_PRA89,Leonhardt_RPP03}, transforming the
  bosonic optical-mode field operators
\begin{align}
  \left(\!\! \begin{array}{c}
          \hat a 
          \\
          \hat b 
        \end{array}\!\right)
  = \hat B(\theta) \; 
  \left(\begin{array}{c}
          \hat a_{\rm in} 
          \\
          \hat b_{\rm in}
        \end{array}\!\right) \; \hat B^\dagger (\theta) \; ,
 \label{eq:U_on_modes}
\end{align}
is given by} 
\begin{equation}\label{eq:U_M}
\hat B (\theta) = \exp[ \frac{\theta}{2} ( \hat a_{\rm in} \hat b_{\rm in}^\dagger - \hat a_{\rm in}^\dagger \hat b_{\rm in} )]
  \equiv \exp[ - {\rm i} \, \tau \, \hat H_{M} ],
\end{equation}
with the effective hamiltonian
\begin{align}\label{eq:H_M}
  \hat H_{M} ={\rm i} \frac{\pi}{2} \left( \hat a_{\rm in} \hat b_{\rm in}^\dagger - \hat a_{\rm in}^\dagger \hat b_{\rm in} \right)
  = \frac{\pi}{2} \left( \hat x_{a} \hat p_{b} -\hat p_{a} \hat x_{b}  \right) \; .
\end{align}

Our choices (together with the fact that the reflection amplitude is~$r = \sin(\frac{\pi}{2}{\tau})$,
and~$t = \cos(\frac{\pi}{2}{\tau})$ the transmission amplitude) allow us to choose the range
$\tau \in [0,1]$ which parameterises the behavior ranging from no mode-mixing at~$\tau=0$ (complete
transparency) via all intermediate values, where the modes are being mixed, to eventually no mixing
at~$\tau=1$, due to total reflection at the beam splitter.

In an implementation of the variable beam splitter as a two-mode fiber coupler, $\tau$ parametrizes
the coupling length, implemented as a Mach-Zehnder interferometer it parametrizes changes of the
phase shifter angle.

The Wigner transform of the associated von Neumann equation gives the evolution equation of
the two-mode Wigner distribution~$W_{ab} = W_{ab}(x_a,p_a,x_b,p_b,\tau)$
\begin{align}\label{eq:vNeumann_M_TrB_0}
\frac{\partial {\cal W}[\hat \rho]}{\partial \tau} = {2\pi\hbar}
  \frac{\partial W_{ab}}{\partial \tau} = {\cal W}\left[ \frac{\pi}{2 {\rm i}\hbar} [ \left( \hat x_{a} \hat p_{b}
  - \hat p_{a} \hat x_{b}  \right), \hat \rho_{ab}] \right].
\end{align}

\subsection{Wigner Current for Mixed Modes}

Let us assume that we choose mode~$a$ as auxil\-ia\-ry, to be traced out of the
state~$\hat \rho_{ab}$. Then the Wigner transform of the von Neumann equation~(\ref{eq:vNeumann_M_TrB_0})
for the partial trace density matrix~$\hat \rho_b$, occupying the remaining mode~$b$, is
\begin{flalign}\label{eq:vNeumann_M_TrB_2}
  2\pi\hbar  \frac{\partial W_b}{\partial \tau} = 
  \frac{\pi}{2} {\cal W} \left[ \frac{1}{{\rm i}\hbar}  {\rm Tr}_a
  \{ [ \hat x_{a} \hat p_{b} , \hat \rho_{ab} ] - [ \hat p_{a} \hat x_{b} , \hat \rho_{ab} ] \}  \right]
\end{flalign}
and hence, using the cyclicity of the trace and Eqns.~(\ref{eq:WignerDistribution})
and~(\ref{eq:moyal_motion}),
\begin{align}\label{eq:Moyal_M_TrB_}
\frac{\partial W_b}{\partial \tau} =
  \frac{\pi}{2} \Big( \{\!\!\{  p_{b} , {\rm Tr}_a \{  x_{a} W_{ab} \} \}\!\!\} -
  \{\!\!\{   x_{b} , {\rm Tr}_a \{  p_{a} W_{ab} \} \}\!\!\} \Big) \; ,
\end{align}
where~$ \{\!\!\{ p_b, {\rm Tr}_a \{ x_{a} \sbullet[.7] \} \}\!\!\}$
and~$ \{\!\!\{ -x_b, {\rm Tr}_a \{ p_{a} \sbullet[.7] \}\}\!\!\}$ are the `effec\-tive Moyal brackets'
taking $W_{ab}$ as an argument, see Eq.~(\ref{eq:moyal_motion}).

Applying Eq.~(\ref{EqMoyalBraket}) leads to 
\begin{flalign}
  \label{eq:Moyal_M_TrA_}
  \frac{\partial W_b 
  }{\partial \tau} =  \; -\frac{\pi}{2} & \Big[   x_{b}  \overleftarrow{\frac{\partial}{\partial x_b}} \overrightarrow{\frac{\partial}{\partial p_b}}
  {\rm Tr}_a \{  p_{a} W_{ab} \}
  \Big. \notag \\
     + & \Big. \;\; p_{b}  \overleftarrow{\frac{\partial}{\partial p_b}} \overrightarrow{\frac{\partial}{\partial x_b}}
    {\rm Tr}_a \{  x_{a} W_{ab} \}
    \Big] \; .
\end{flalign}
According to continuity equation~(\ref{eq:continuity}) this is rewritten
as~\cite{Oliva_Kerr_18,Oliva_PhysA17,Ole_23photonAddition}
\begin{flalign}
  \label{eq:Moyal_M_TrB_continuity}
 \frac{\partial W_b}{\partial \tau} = -  \frac{\pi}{2} \! \left(\!\begin{array}{c} \frac{\partial}{\partial x_{b}}
          \\
 \frac{\partial}{\partial p_{b}} \end{array}\!\right) \!\! \cdot \!\!
  \left(\begin{array}{c} {\rm Tr}_a \{ x_{a} W_{ab} \}
          \\
          {\rm Tr}_a \{ p_{a} W_{ab} \} \end{array}\right) \! = - \VEC{\nabla}_b  \cdot 
   \VEC{J}_b \; ,
\end{flalign}
where the Wigner current of mode~$b$, conditioned on the state of mode~$a$, is
\begin{flalign}
  \label{eq:_J_B_explicit}
  \VEC{J}_b (x_b, p_b, \tau) = + \frac{\pi}{2}
  \left(\begin{array}{c} {\rm Tr}_a \{ x_{a} W_{ab} \} (x_b, p_b, \tau)
          \\
          {\rm Tr}_a \{ p_{a} W_{ab} \} (x_b, p_b, \tau)  
        \end{array}\right)  \; .
    \end{flalign}

An analogous calculation yields Wigner current,~$\VEC{J}_a$, of
mode~$a$, conditioned on the state of mode~$b$, as 
\begin{flalign}
  \label{eq:_J_A_explicit}
  \VEC{J}_a (x_a, p_a, \tau) = - \frac{\pi}{2}
  \left(\begin{array}{c} {\rm Tr}_b \{ x_{b} W_{ab} \} (x_a, p_a, \tau)
          \\
          {\rm Tr}_b \{ p_{b} W_{ab} \} (x_a, p_a, \tau)  
        \end{array}\right)  \; .
\end{flalign}

\begin{figure}[t] \centering
  \includegraphics[width=3.8cm,height=3.2cm]{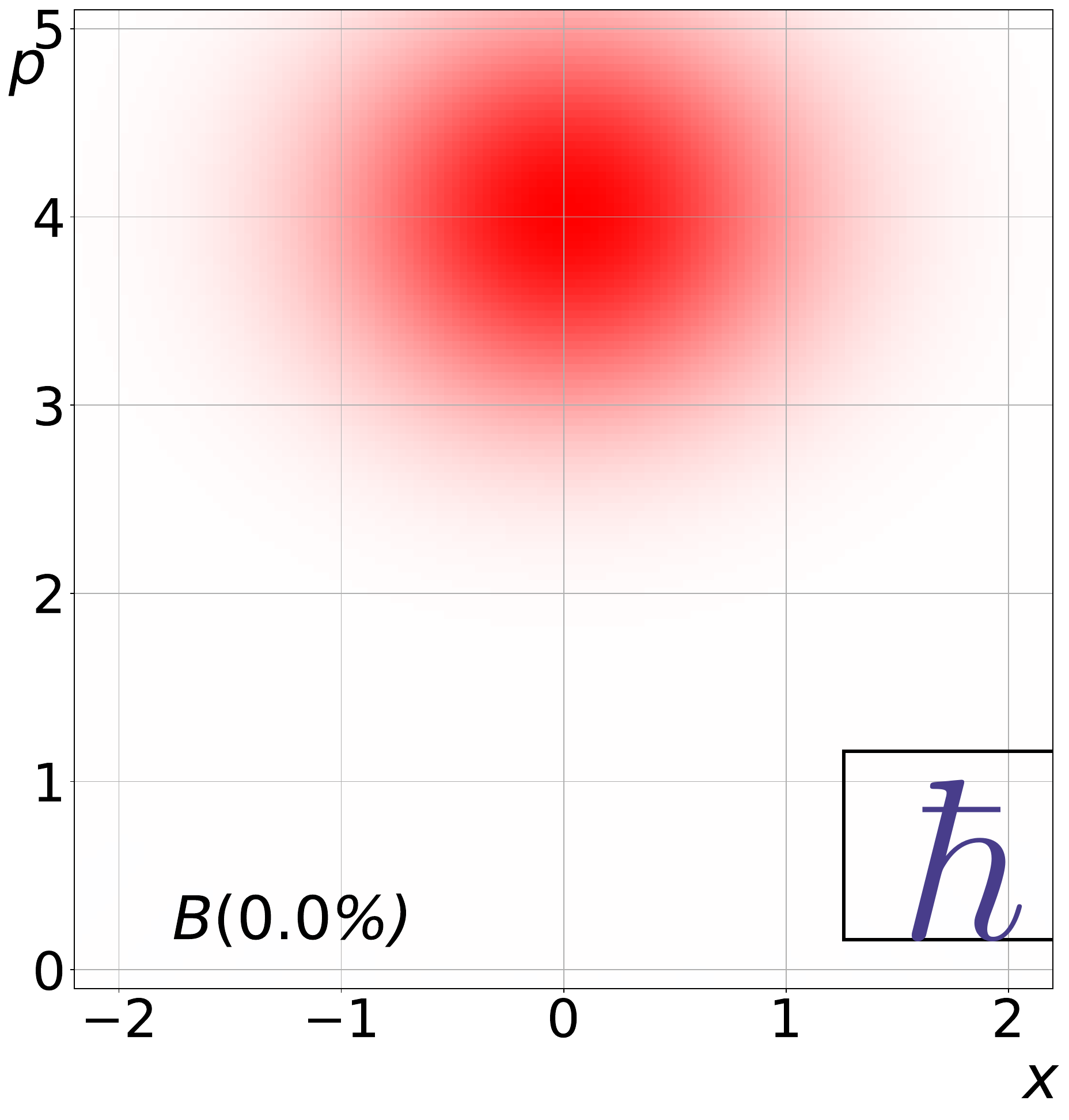}
  \includegraphics[width=4.5cm,height=3.4cm]{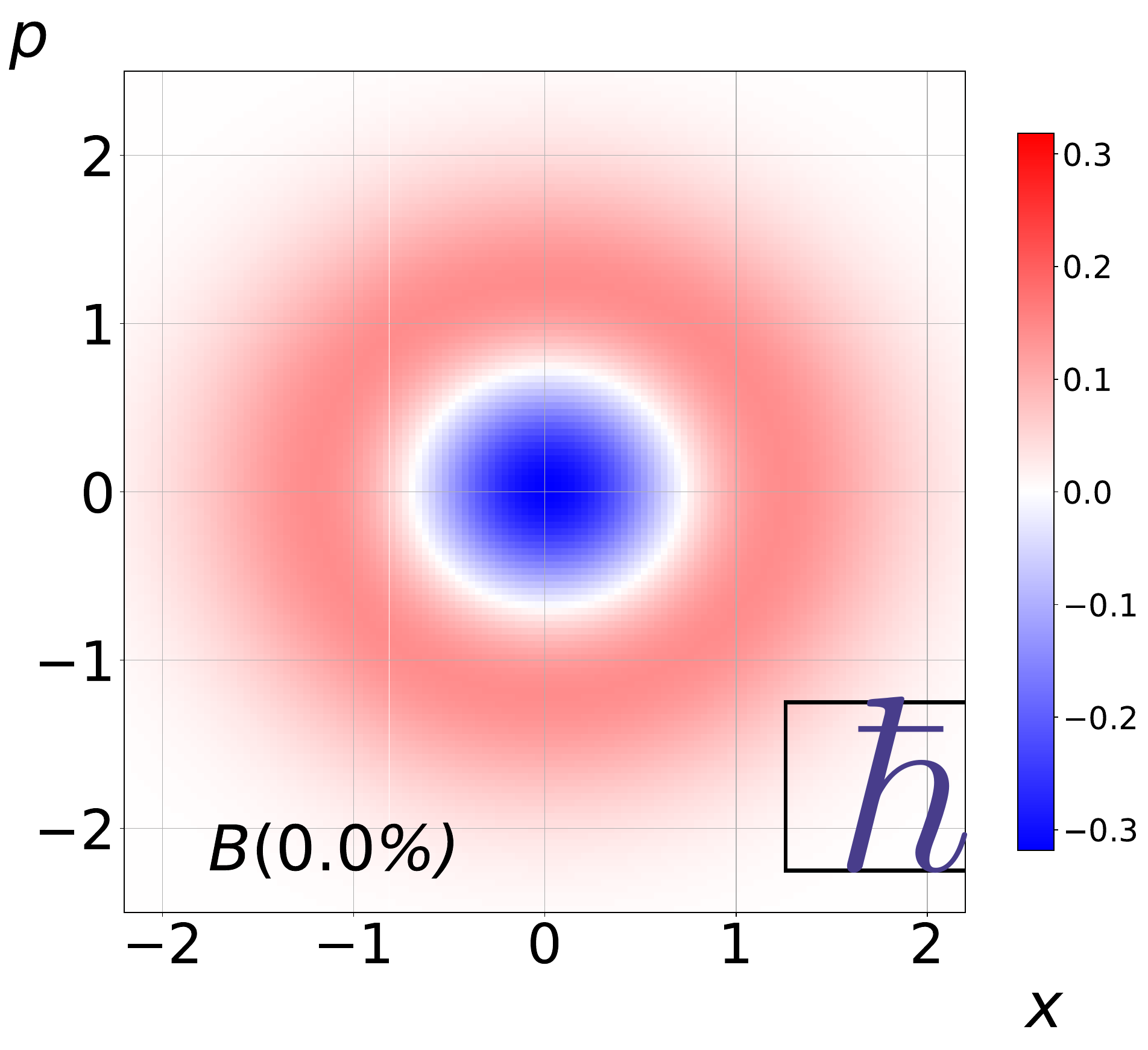}  
\\
  \includegraphics[width=3.8cm,height=3.3cm]{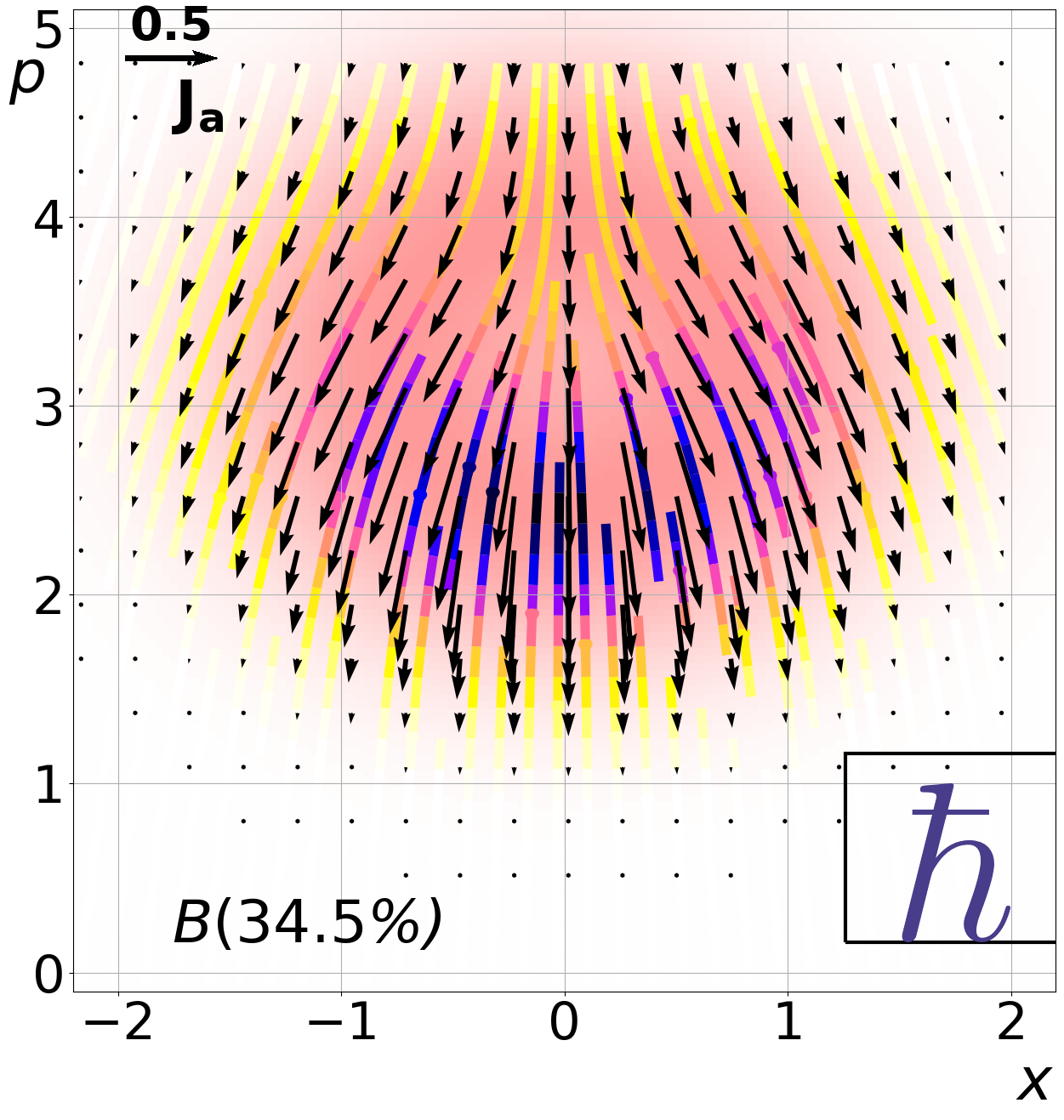}
  \includegraphics[width=4.5cm,height=3.4cm]{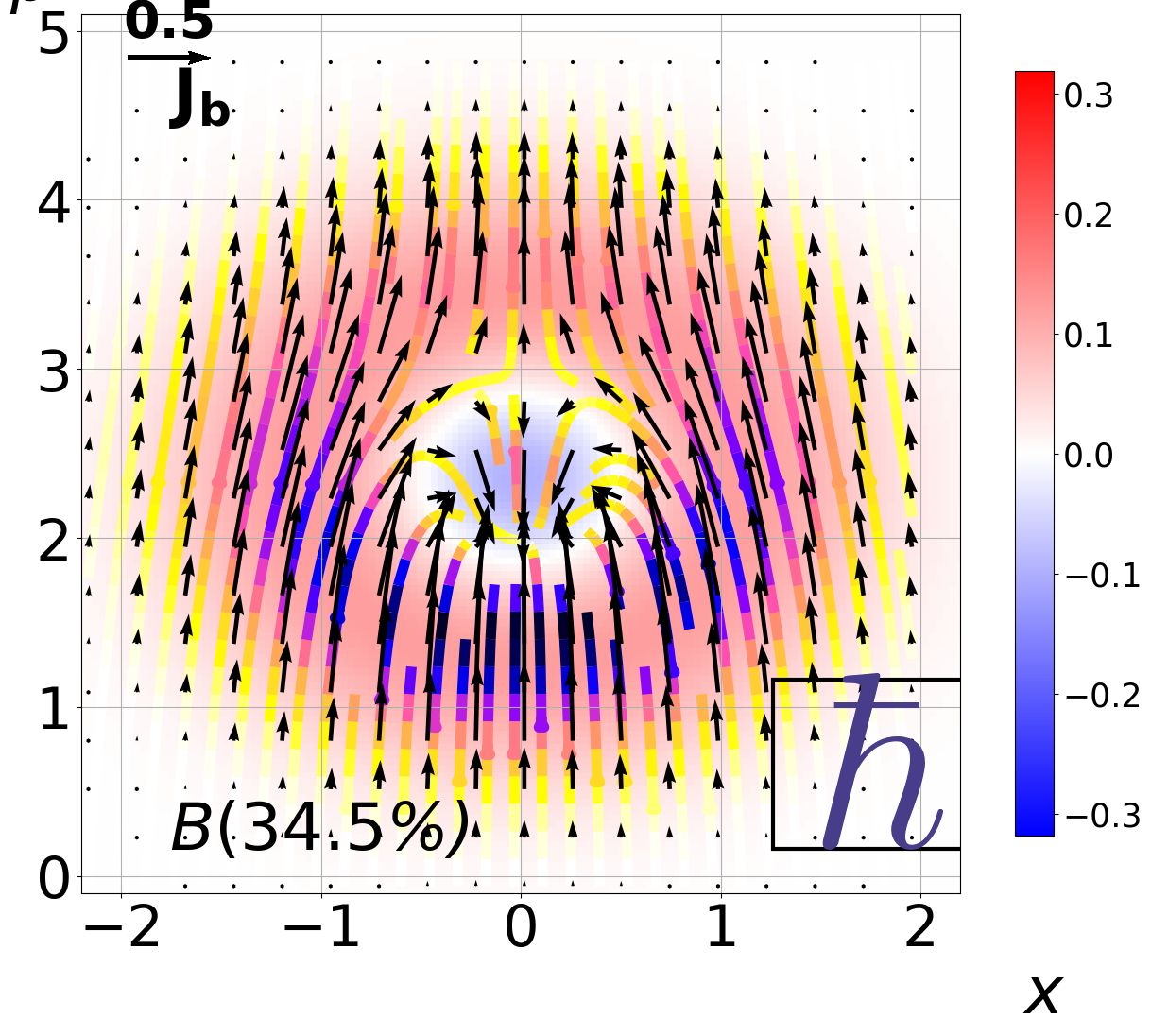}  
  \caption{Wigner distributions~$W_{a\setminus b}$ of modes $a$ (left panels) and $b$ (right
    panels), the top row shows the initial states, the bottom row evolved states together with their
    vector fields~$\VEC{J}_{a\setminus b}$ as overlays. As a guide to the eyes, the $\VEC{J}$-fields
    are integrated yielding white-yellow-blue (coloring according to $|{\VEC J}|$) field lines which
    represent momentary snapshots of the $\VEC{J}$-fields. Thick black frames surround an area of
    size \setlength\fboxsep{0pt}$\!\boxed \hbar$ in \ps. $B(34.5\%)$ represents mixing at a beam
    splitter with 34.5\% reflectivity. The initial states are a coherent state and a single-photon
    Fock state: $|\psi_a \rangle {} |\psi_b \rangle = |\alpha = 4$i$/\sqrt{2} \rangle\otimes |1\rangle$.
  \label{fig:SqeFock1}}
\end{figure}

\subsection{Contrast with Classical Mechanics\label{sec:ClassicalContrast}}

We would like to stress that, despite the existence of continuity equations, currents and phase
portraits in quantum \ps
formulations~\cite{Ole_PRL13,Kakofengitis_EPJP17,Skodje_PRA89,Donoso_PRL01,Bauke_2011arXiv1101.2683B,Veronez_JPA13,Valtierra_PRA20},
in general, trajectories in quantum \ps do \emph{not exist}~\cite{Oliva_PhysA17}.  We neither
discuss paths (such as Feynman's, which do not have to follow equations of motion) nor
center-of-mass trajectories as famously discussed by Heisenberg in his 1927 paper on interpretations
of \emph{observed} particle trajectories~\cite{Heisenberg_ZfP27}.

We remind the reader that, trajectories are integral curves that obey the equations of motion
\emph{and} describe the transport of probability. In quantum mechanics these do not generally exist
since \ps velocities would develop singularities~\cite{Oliva_PhysA17}.

Although trajectories do not exist, we can determine \ps field lines, the integral curves
of~$\VEC J$ at a fixed time, see lower panels of Fig.~\ref{fig:SqeFock1} (and also
Fig.~\ref{fig:SquSqu}).  Specifically, in classical mechanics' \ps settings such curves are also
known as phase curves and their collection as phase portraits, which is why we introduce the term
`quantum phase portraits' denoting such collections of field lines.

\begin{figure}[t] \centering
  \includegraphics[width=3.8cm,height=3.2cm]{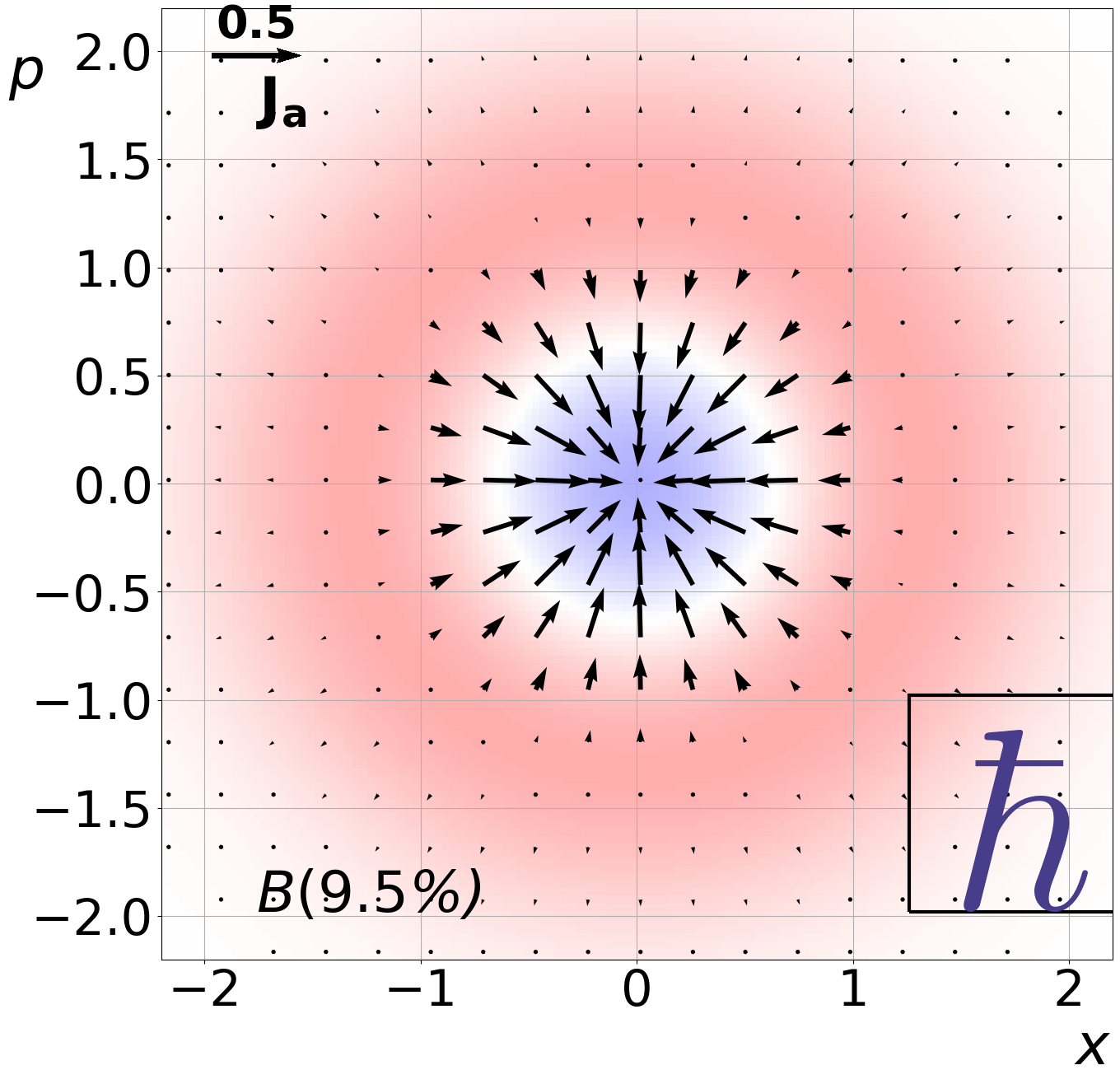}
    \setlength{\fboxsep}{0pt}
    \begin{picture}(0,0) \put(-79,69){\fcolorbox{brown}{white}{\includegraphics[height=1.0cm
        ]{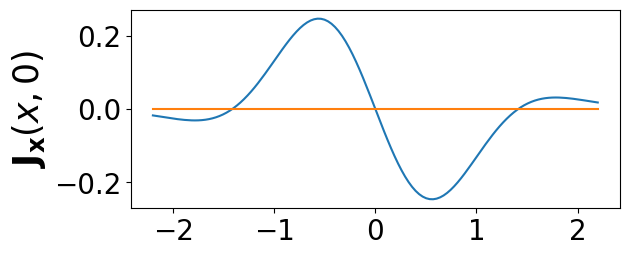}}}
  \end{picture}
  \includegraphics[width=4.5cm,height=3.4cm]{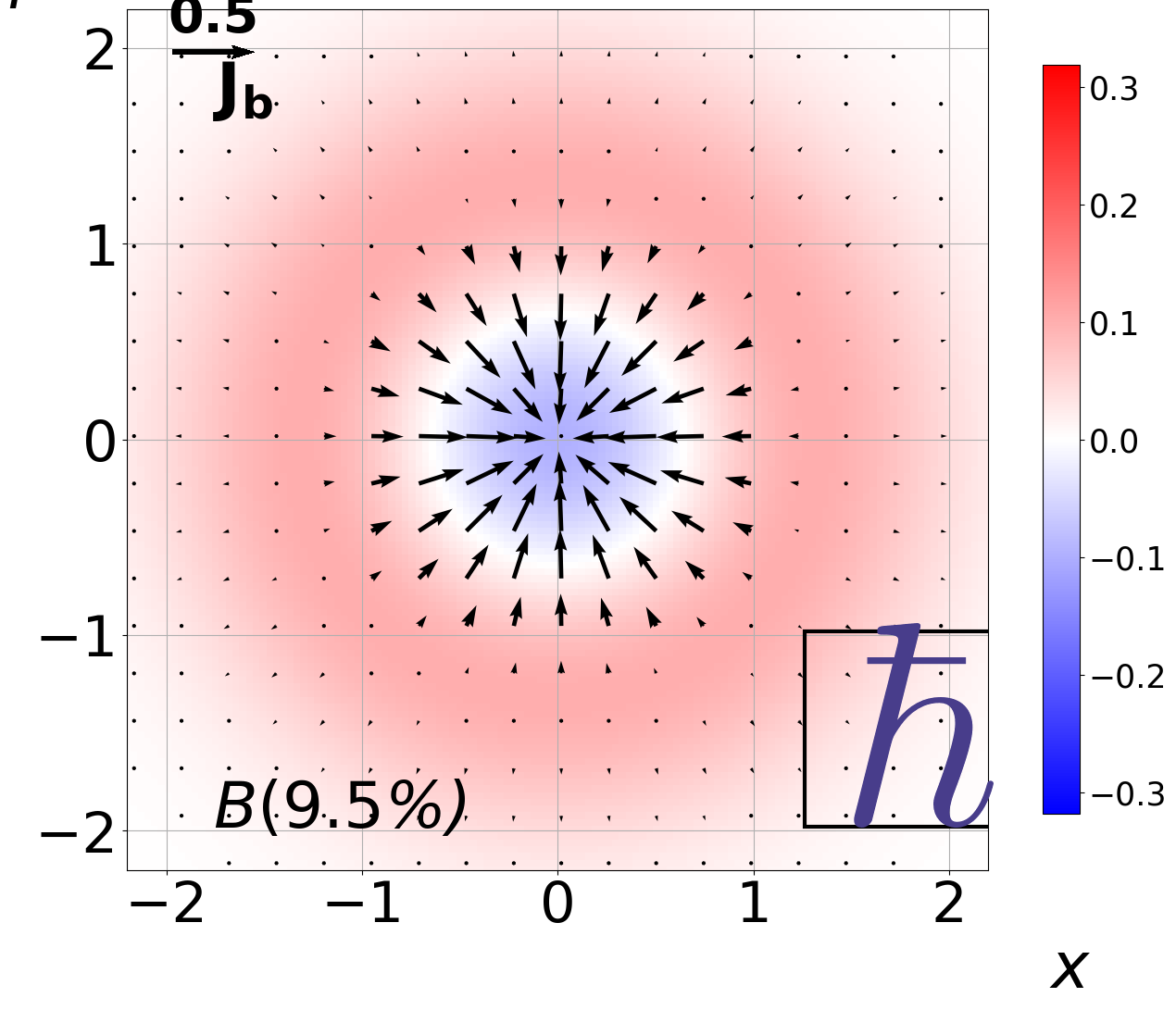}
    \setlength{\fboxsep}{0pt}
    \begin{picture}(0,0) \put(34,79){\fcolorbox{blue}{white}{\includegraphics[height=1.0cm
        ]{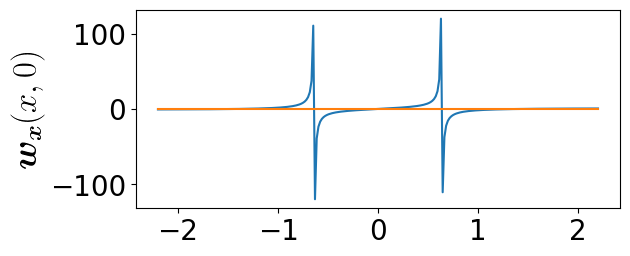}}}
  \end{picture}
  \\
  \includegraphics[width=3.8cm,height=3.2cm]{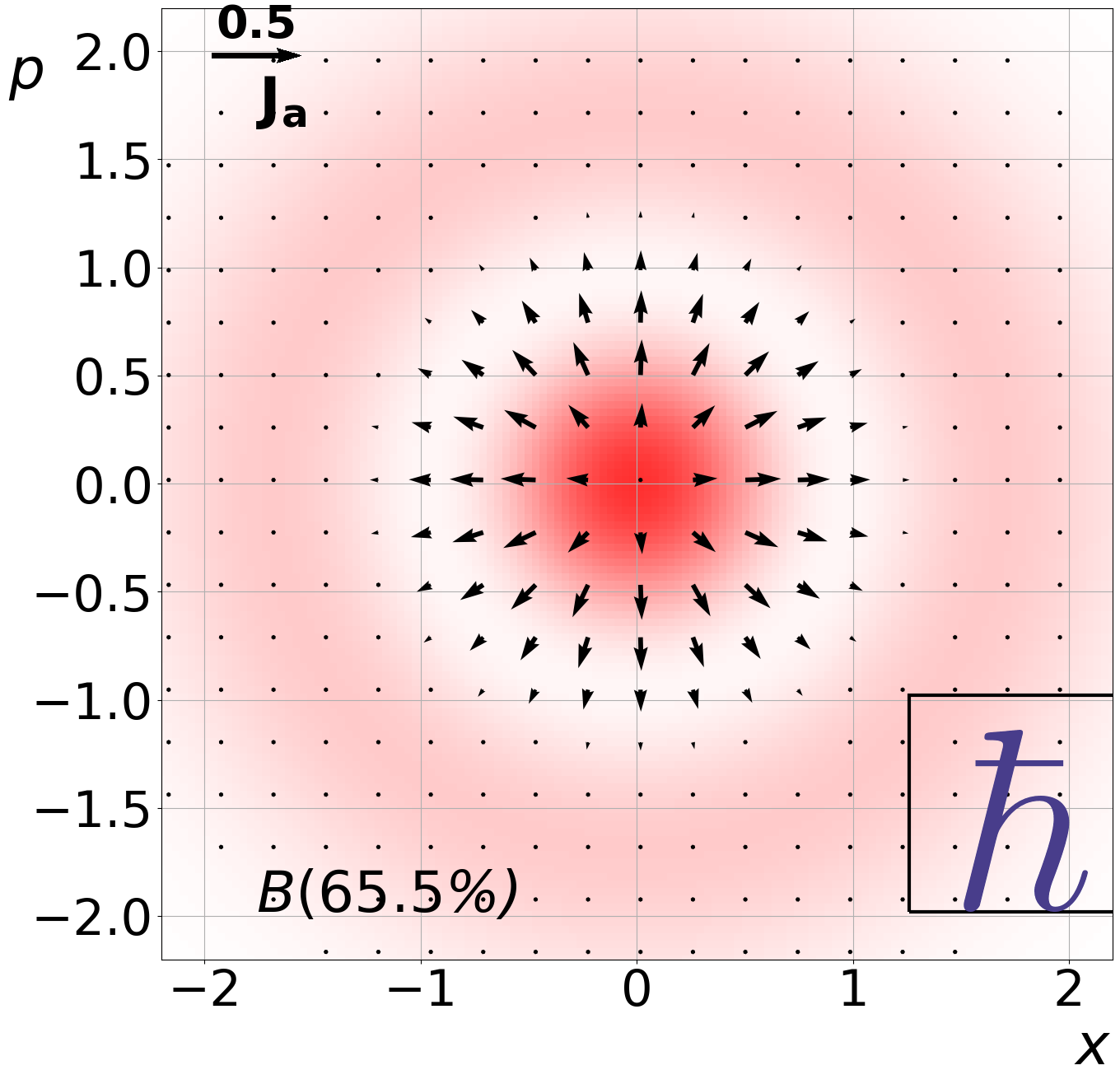}
    \setlength{\fboxsep}{0pt}
  \begin{picture}(0,0) \put(-79,69){\fcolorbox{brown}{white}{\includegraphics[height=1.0cm
        ]{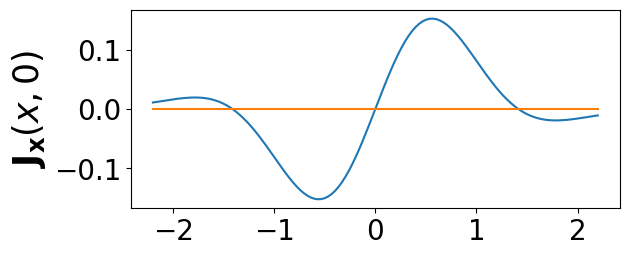}}}
  \end{picture}
  \includegraphics[width=4.5cm,height=3.4cm]{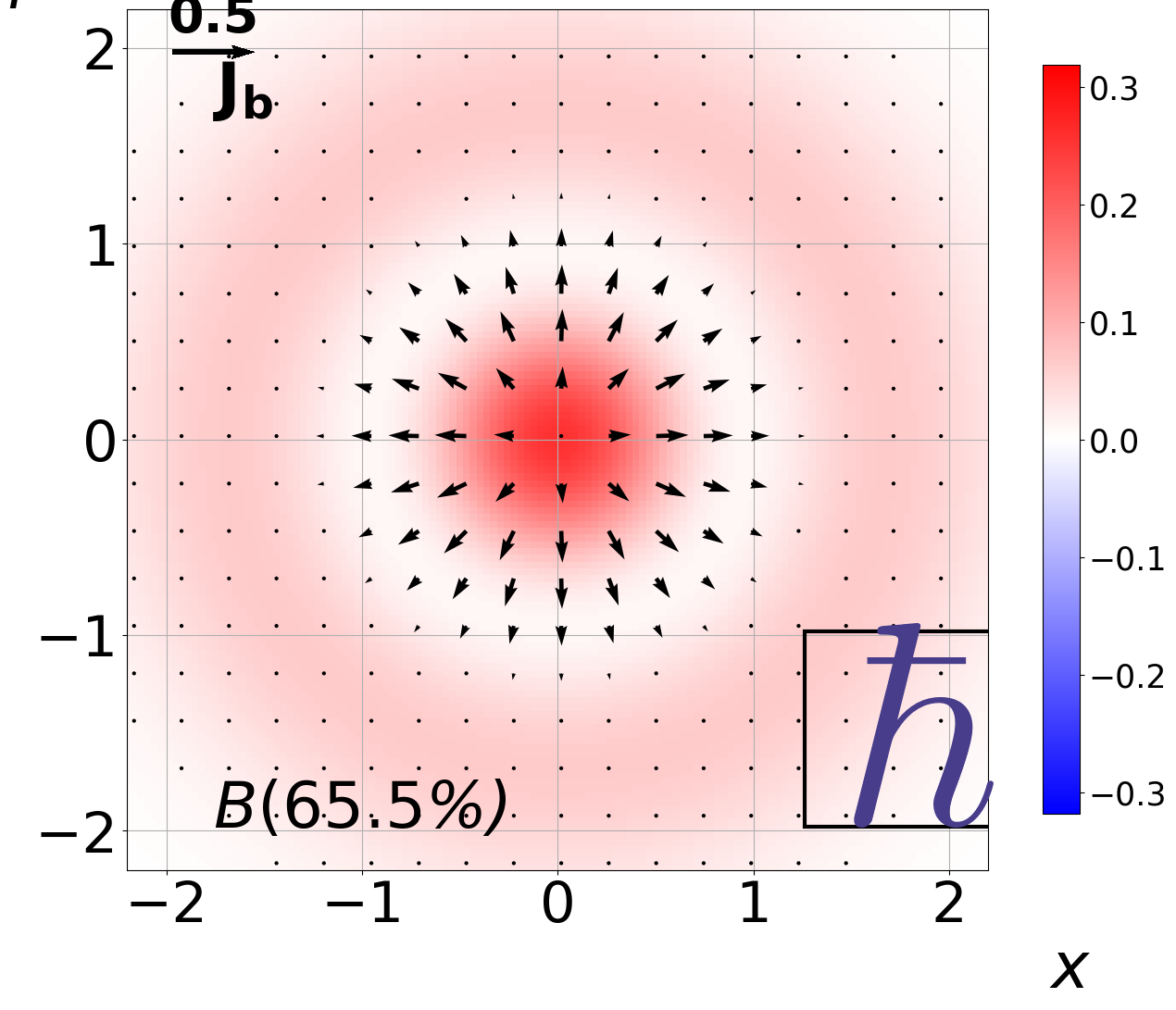}
    \setlength{\fboxsep}{0pt}
    \begin{picture}(0,0) \put(34,79){\fcolorbox{blue}{white}{\includegraphics[height=1.0cm
        ]{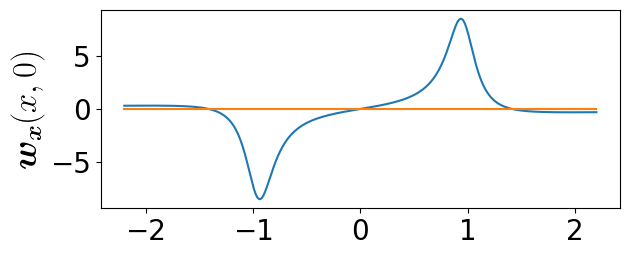}}}
  \end{picture}
  \\
  \includegraphics[width=3.8cm,height=3.2cm]{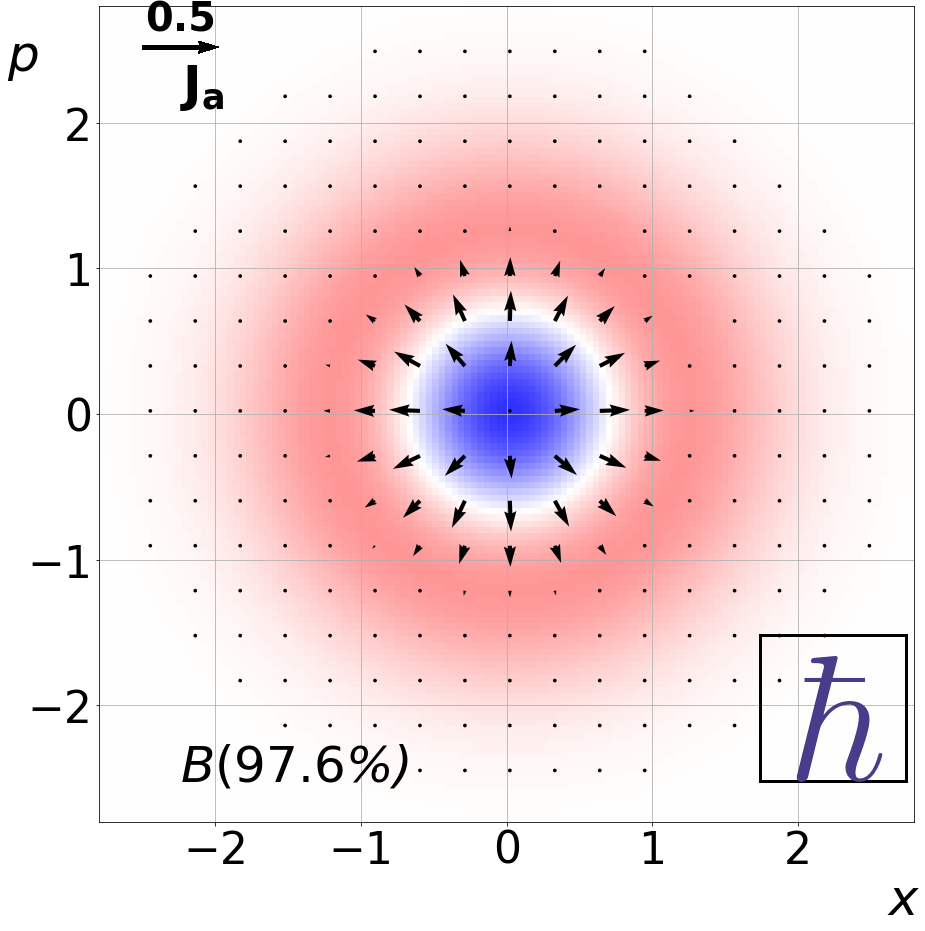}
    \setlength{\fboxsep}{0pt}
  \begin{picture}(0,0) \put(-79,69){\fcolorbox{brown}{white}{\includegraphics[height=1.0cm
        ]{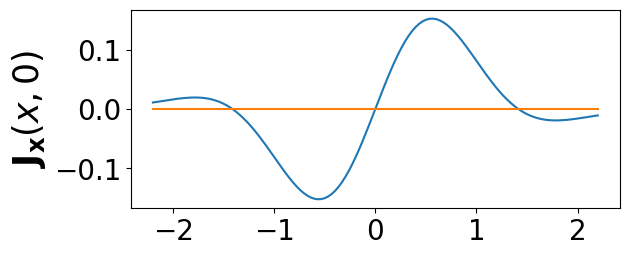}}}
  \end{picture}
  \includegraphics[width=4.5cm,height=3.4cm]{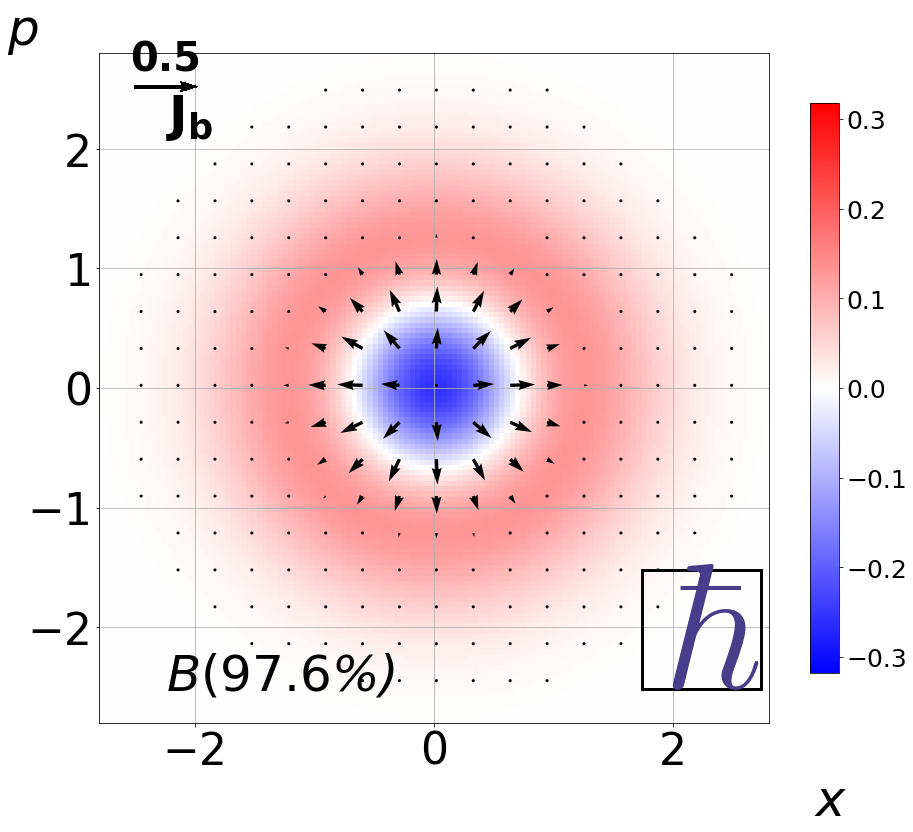}
      \setlength{\fboxsep}{0pt}
    \begin{picture}(0,0) \put(34,79){\fcolorbox{blue}{white}{\includegraphics[height=1.0cm
        ]{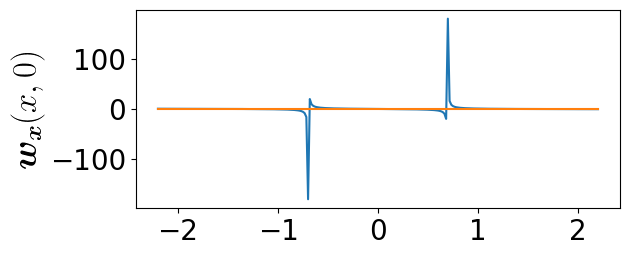}}}
  \end{picture}
  \caption{Layout as in Fig.~\ref{fig:SqeFock1}.  The initial states are two single-photon Fock
    states: $|\psi_a \rangle {} |\psi_b \rangle = |1\rangle\otimes |1\rangle$ (see top right panel
    of Fig.~\ref{fig:SqeFock1}) and modes $a$ and $b$ behave identically.
    Insets in panels on the left column (brown frames)
    show plots of the  $x$-component ${\VEC J}_x(x,0)$ along the $x$-axis.
    Insets in panels on the right column (blue frames)
    show plots of the $x$-component ${\VEC w}_x(x,0)$ along the $x$-axis and that it is singular
    when~$W=0$.
    \label{fig:HOM}}
\end{figure}

\begin{figure}[b] \centering
  \includegraphics[width=3.8cm,height=3.2cm]{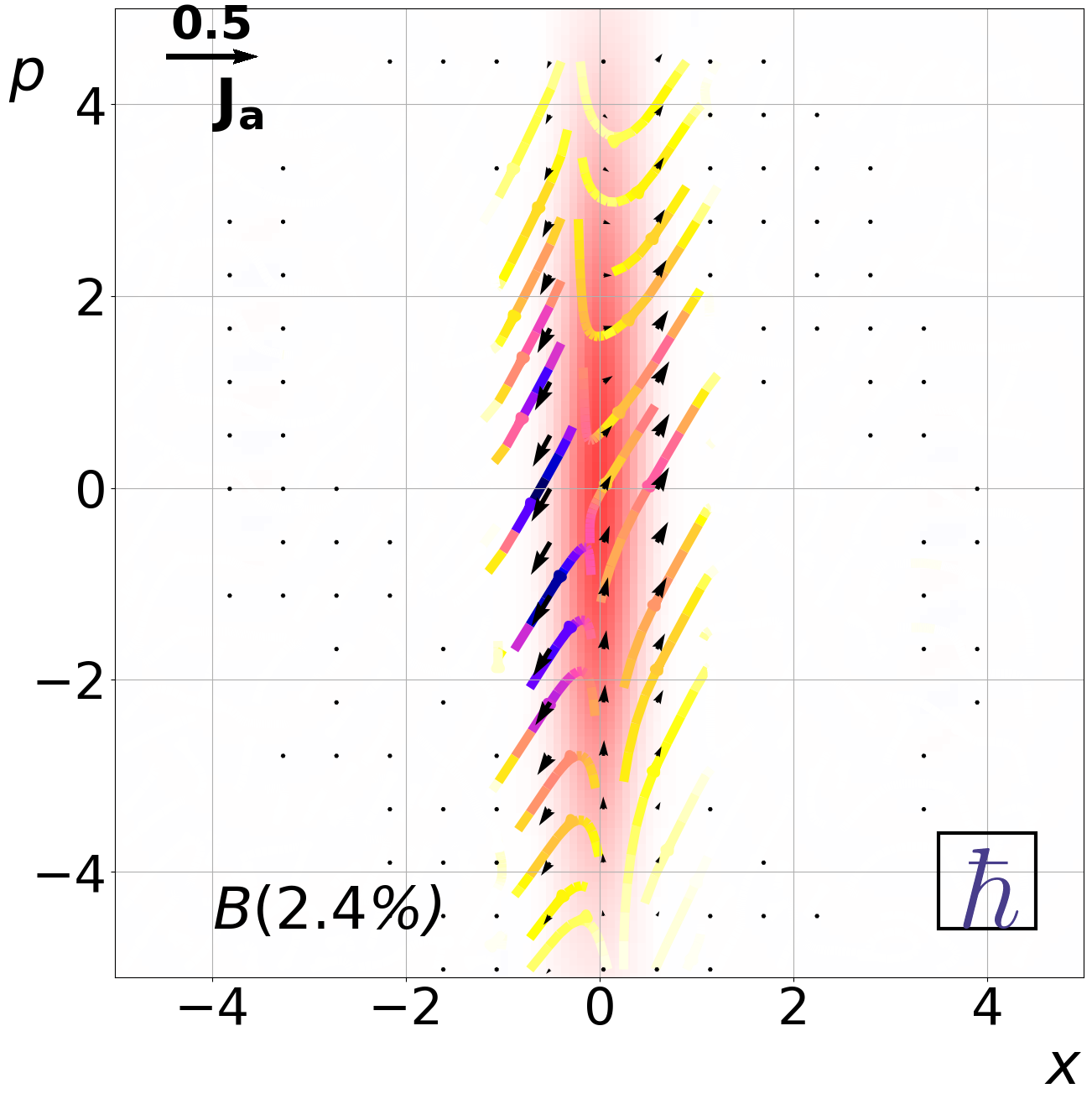}
  \includegraphics[width=4.5cm,height=3.4cm]{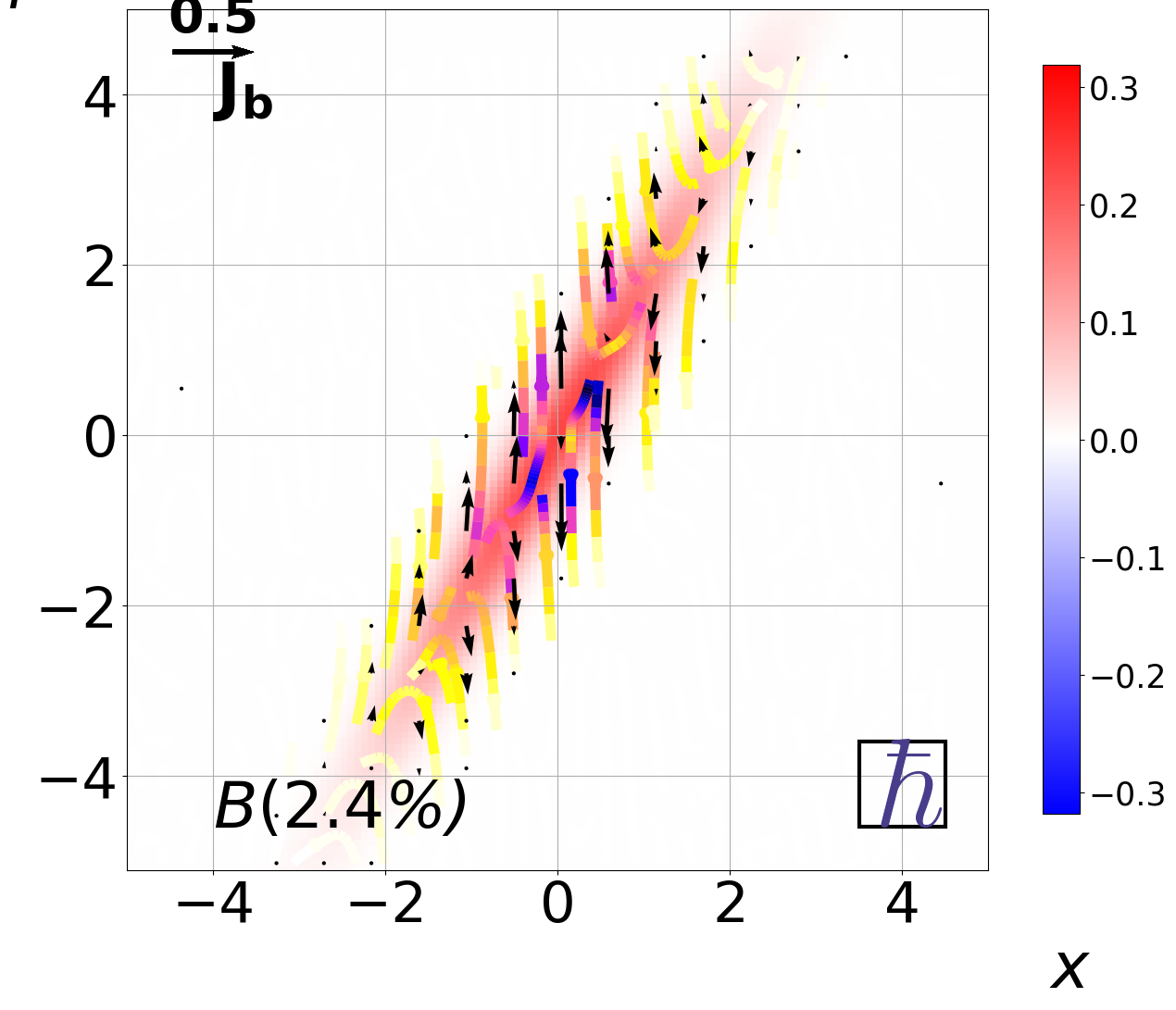}
  \\
  \includegraphics[width=3.8cm,height=3.2cm]{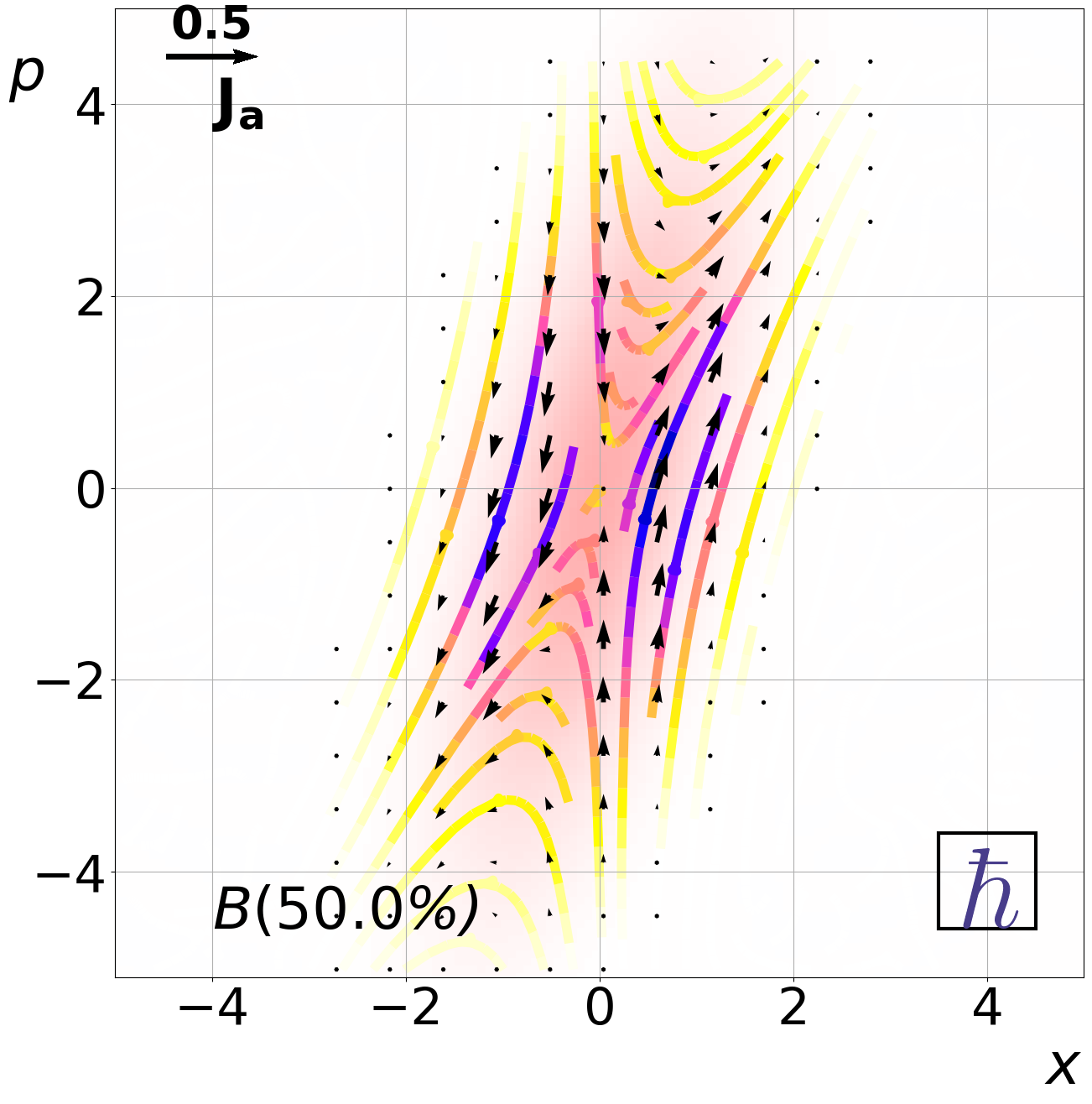}
  \includegraphics[width=4.5cm,height=3.4cm]{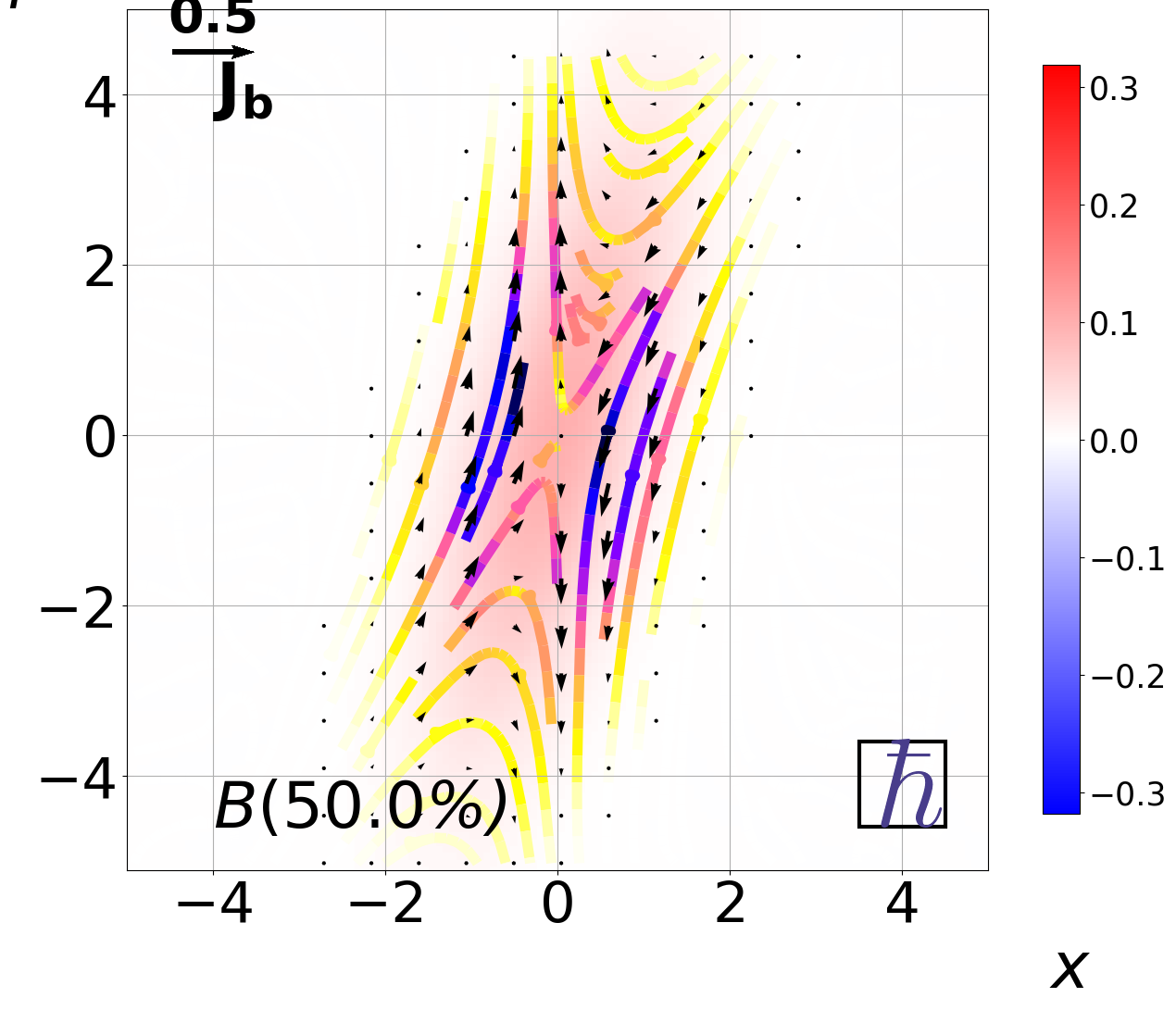}
  \\
  \includegraphics[width=3.8cm,height=3.2cm]{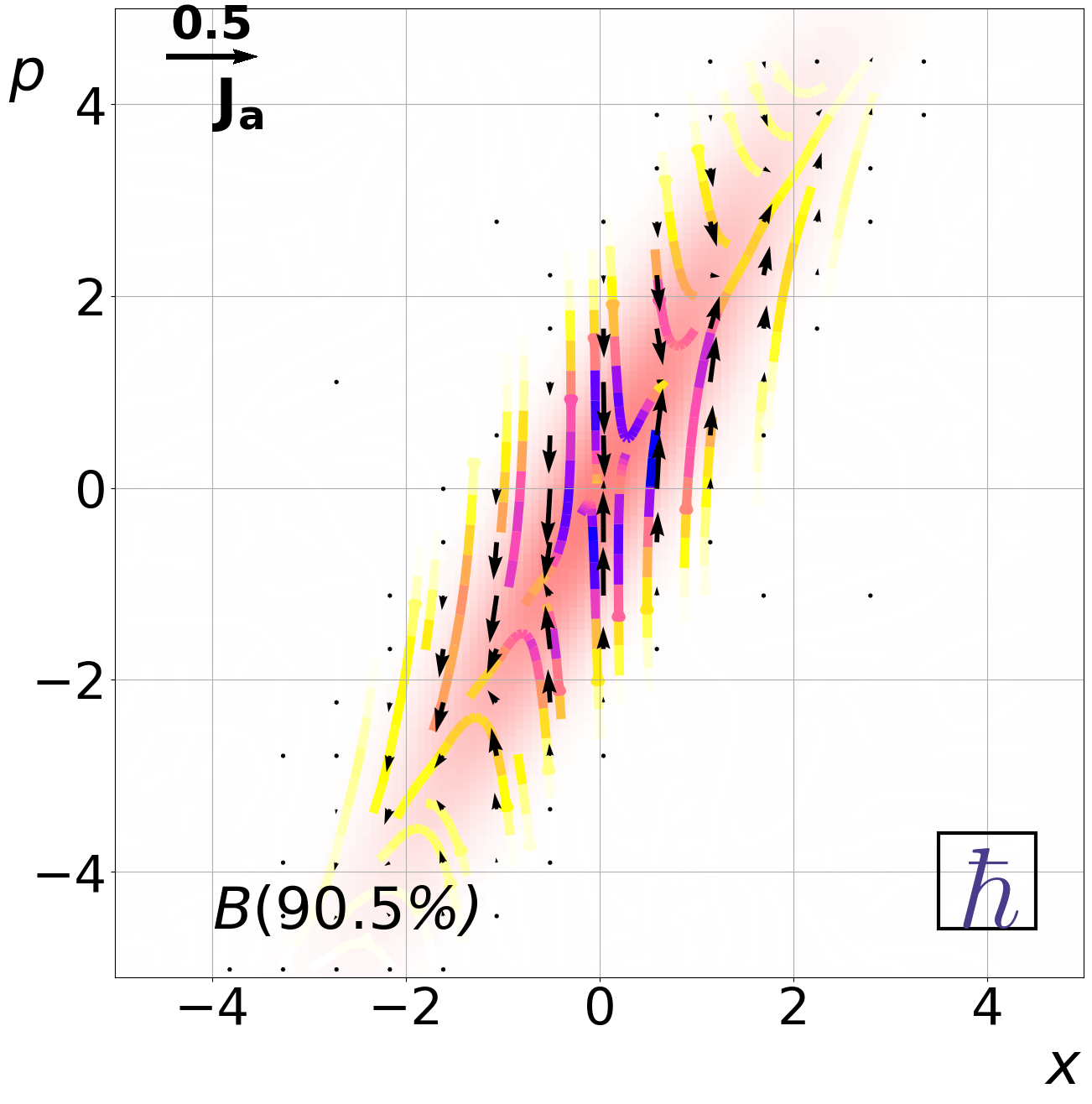}
  \includegraphics[width=4.5cm,height=3.4cm]{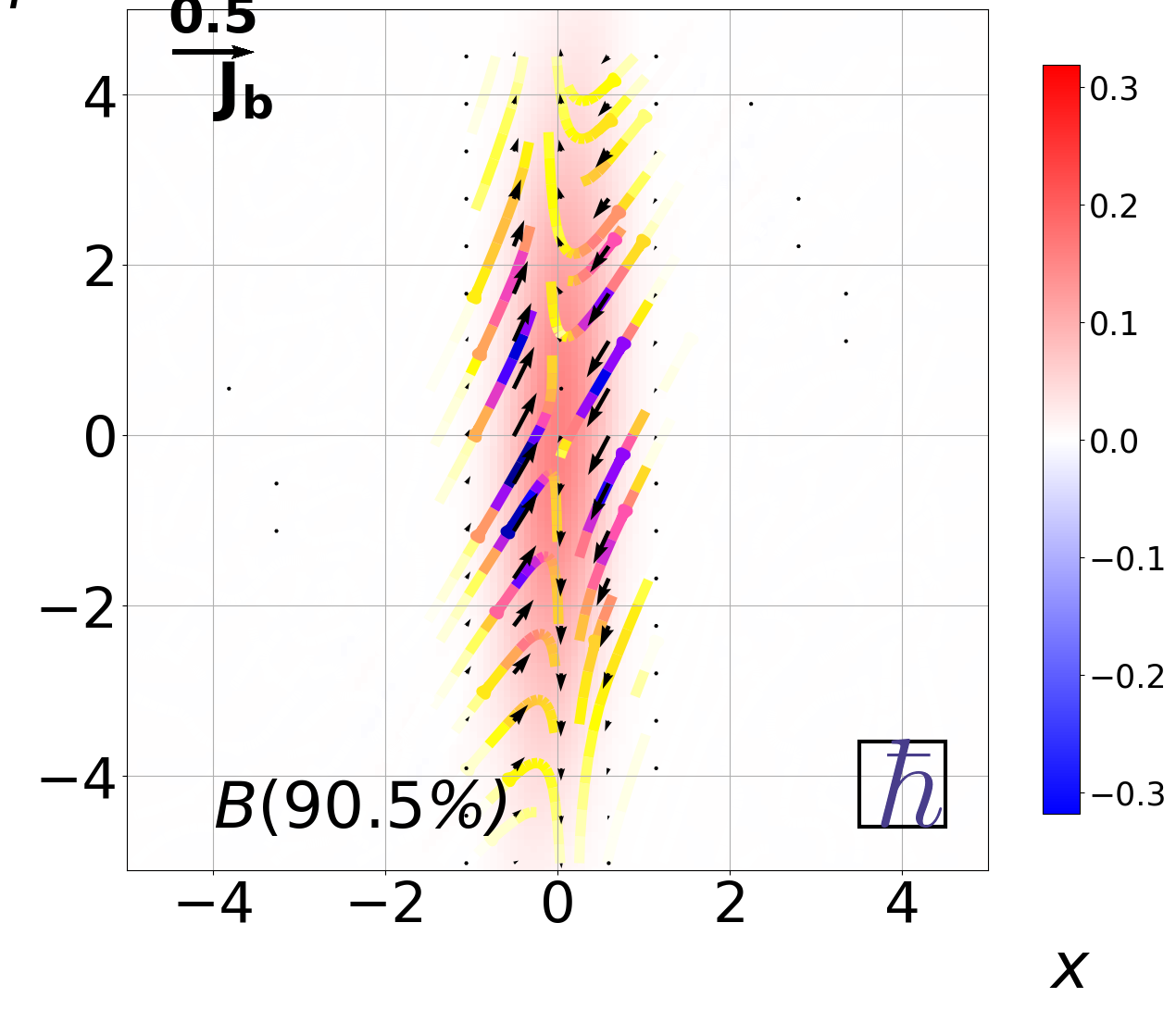}
  \caption{Layout as in Fig.~\ref{fig:SqeFock1}, with two equally strongly squeezed vacua: $|\psi_a \rangle  {}
    |\psi_b \rangle = |z=2,\theta = 0\rangle\otimes |z=2,\theta = -\pi/3 \rangle$
    as  initial states.
    \label{fig:SquSqu}}
\end{figure}

Let us briefly explain why singular volume changes occur in quantum \ps. In classical physics the
current factorizes into density and classical \ps velocity field~\cite{Oliva_PhysA17}
$\VEC{j} = \rho \VEC{v}$, therefore $\VEC{v} = \VEC{j} / \rho $. In quantum dynamics, however,
attempting this approach, namely, defining the quantum \ps velocity field, $\VEC{w} = \VEC{J}/W$,
typically yields singular values somewhere in \ps, since the Moyal bracket contains derivatives of
$W$, which is why zeros of $W$ do not need to coincide with those of~$\VEC{J}$, see
Fig.~\ref{fig:HOM}. Therefore, the degeneracy between zeros of $W$ and $\VEC J$ is typically lifted
and when $W$ is zero $\VEC{w} = \VEC{J} /W $ is singular, see insets in Right Column of
Fig.~\ref{fig:HOM}. Liouville's classical \ps volume conservation is described by the condition
$\VEC{\nabla} \cdot \VEC{v} = 0$, the equivalent quantum expression $\VEC{\nabla} \cdot \VEC{w} $
can assume values with infinite magnitude~\cite{Oliva_PhysA17}.

This makes trajectories approaches flawed, they are based on $ \VEC{w} $, and so develop
singularities and singular changes in \ps volumes when $W=0$; in short, there are no trajectories in
quantum \ps~\cite{Oliva_PhysA17}.

One should avoid using the unfortunate term `quantum Liouville equation', which has caused great
confusion when researchers talk about so-called `quantum characteristics' in \ps and try to
implement trajectories-based code on computers~\cite{Oliva_PhysA17}. Instead, correctly set up and
efficient `exact' numerical codes for the propagation of Wigner's distribution using spectral
methods do exist~\cite{Cabrera_PRA15}, also in the case of nonlinear wave
equations~\cite{Ole_JPAMT23} and nonsepa\-rable hamiltonians~\cite{Ciric_EJPP23}.

\section{Results for Various Field States\label{sec:Examples}}

In this section we consider initially unentangled
states~$|\psi_a \rangle {} |\psi_b \rangle = |\psi_a \rangle \otimes |\psi_b \rangle$ for the
incoming beam splitter modes~$ a_{\rm in}$ and~$b_{\rm in}$, using the following notation:

$N$-photon Fock states are denoted by~$|N\rangle$, degenerate squeezed vacuum states
$\text{exp}[\frac{1}{2} ( \zeta^\ast \hat{a}^2 - \zeta \hat{a}^{\dag 2})]|0\rangle$ by
$|z=|\zeta|, \theta=$arg$(\zeta)\rangle$, and Glauber cohe\-rent states
\mbox{$\text{exp}[\alpha\hat a^\dag - \alpha^*\hat a]$}$|0\rangle$
by~$|\sqrt{2}\alpha = x +$i$p\rangle$.

\begin{figure}[t] \centering
  \includegraphics[width=3.8cm,height=3.2cm]{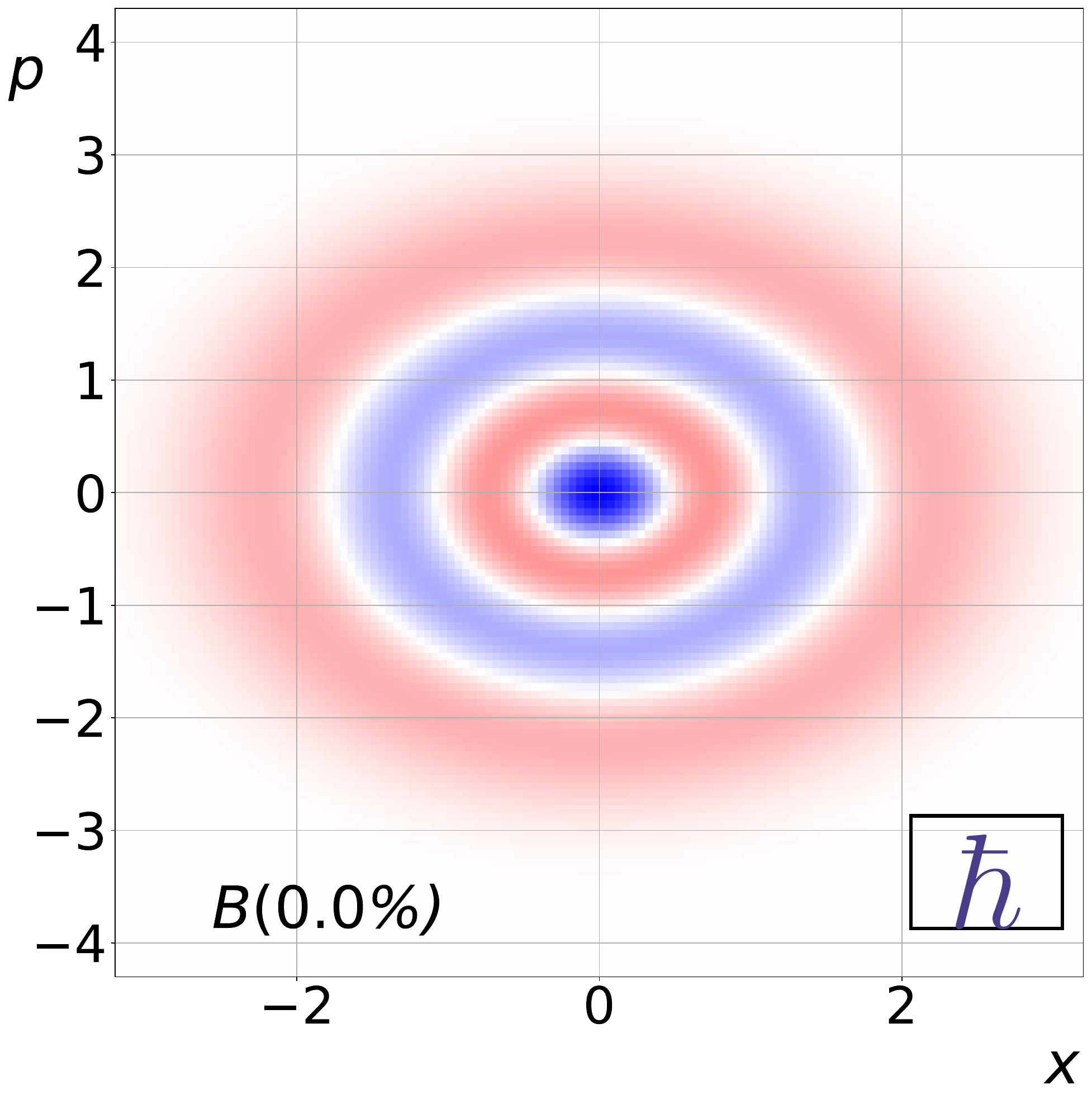}
  \includegraphics[width=4.5cm,height=3.4cm]{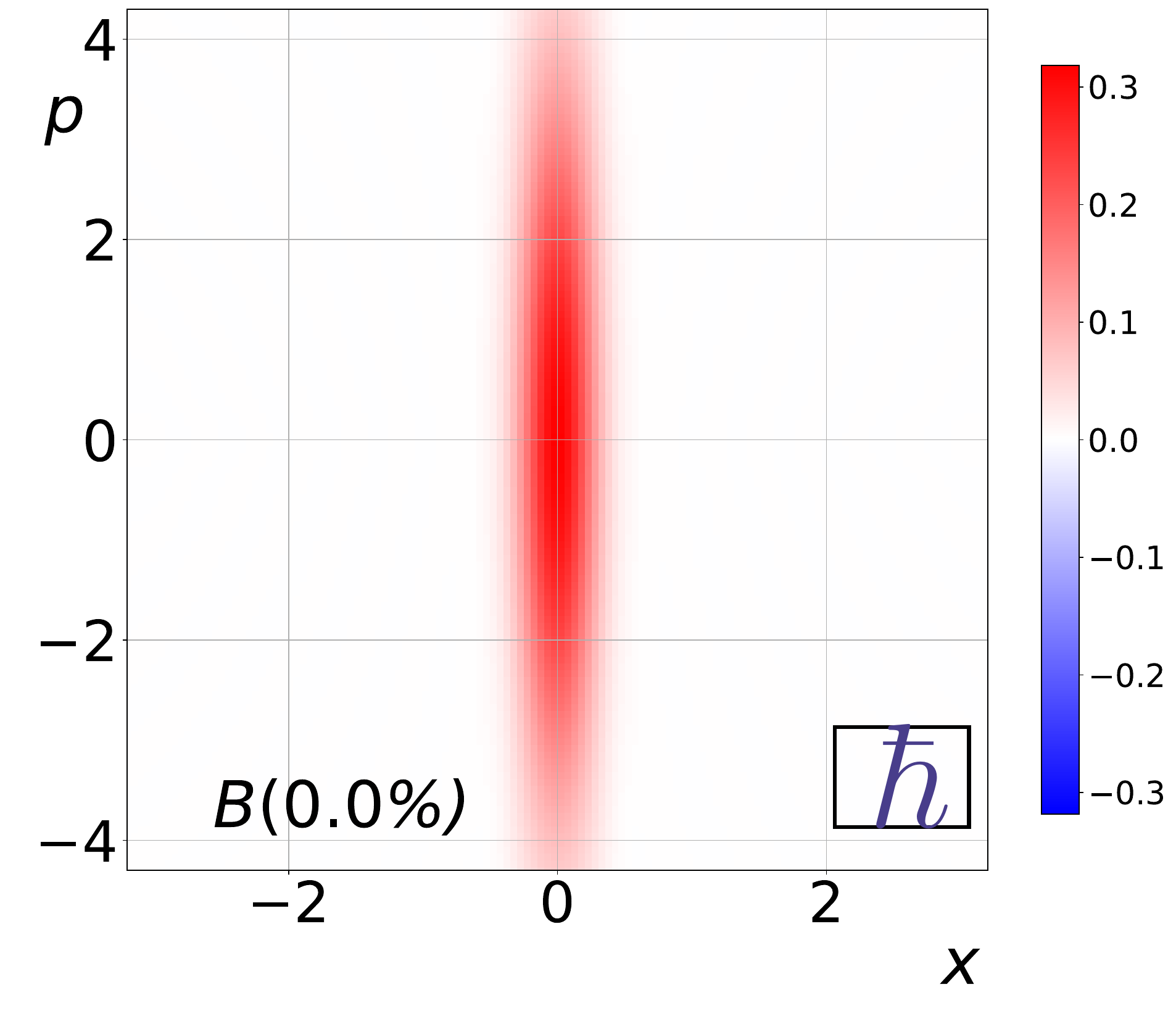}
  \\
  \includegraphics[width=3.8cm,height=3.2cm]{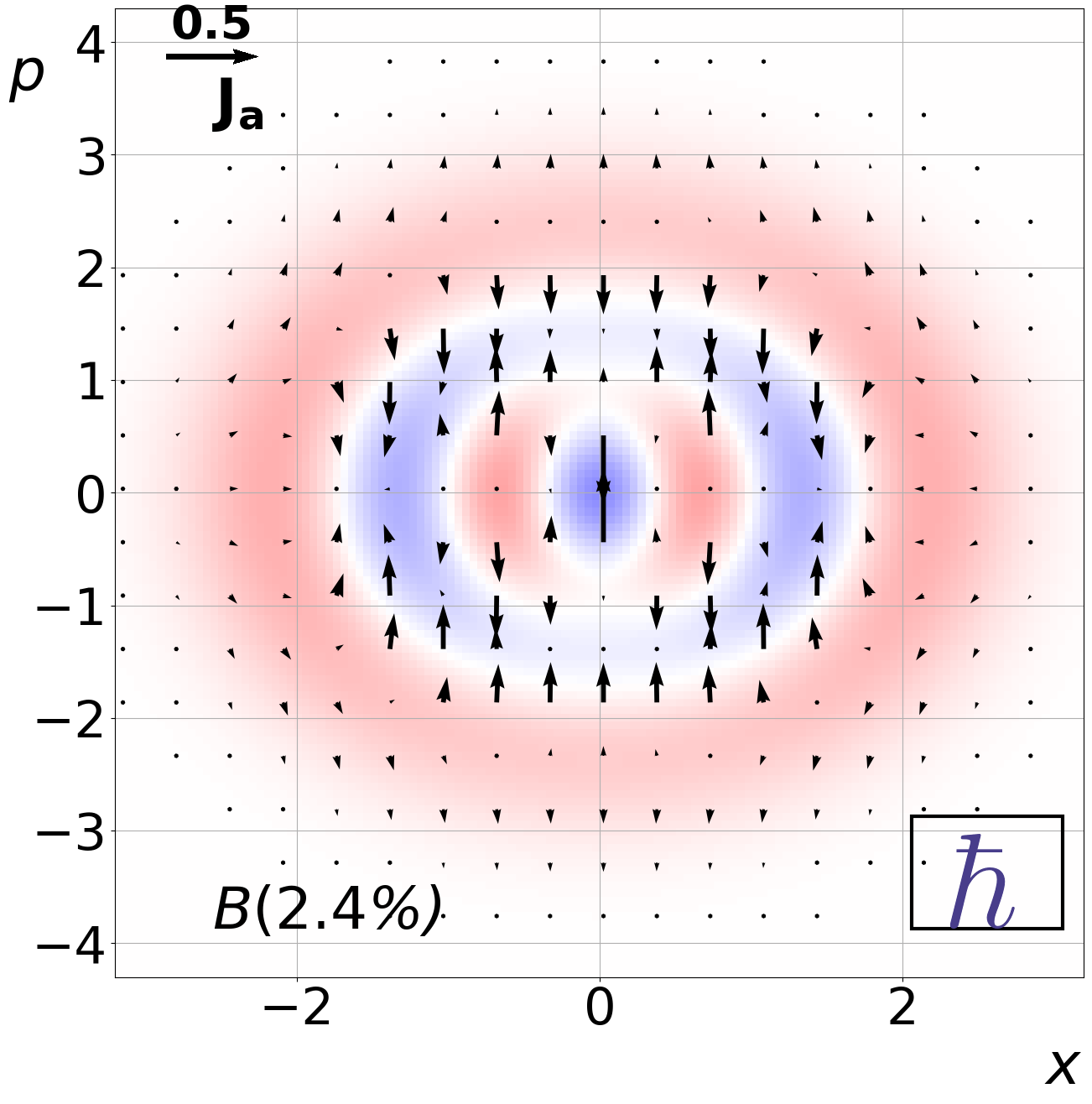}
  \includegraphics[width=4.5cm,height=3.4cm]{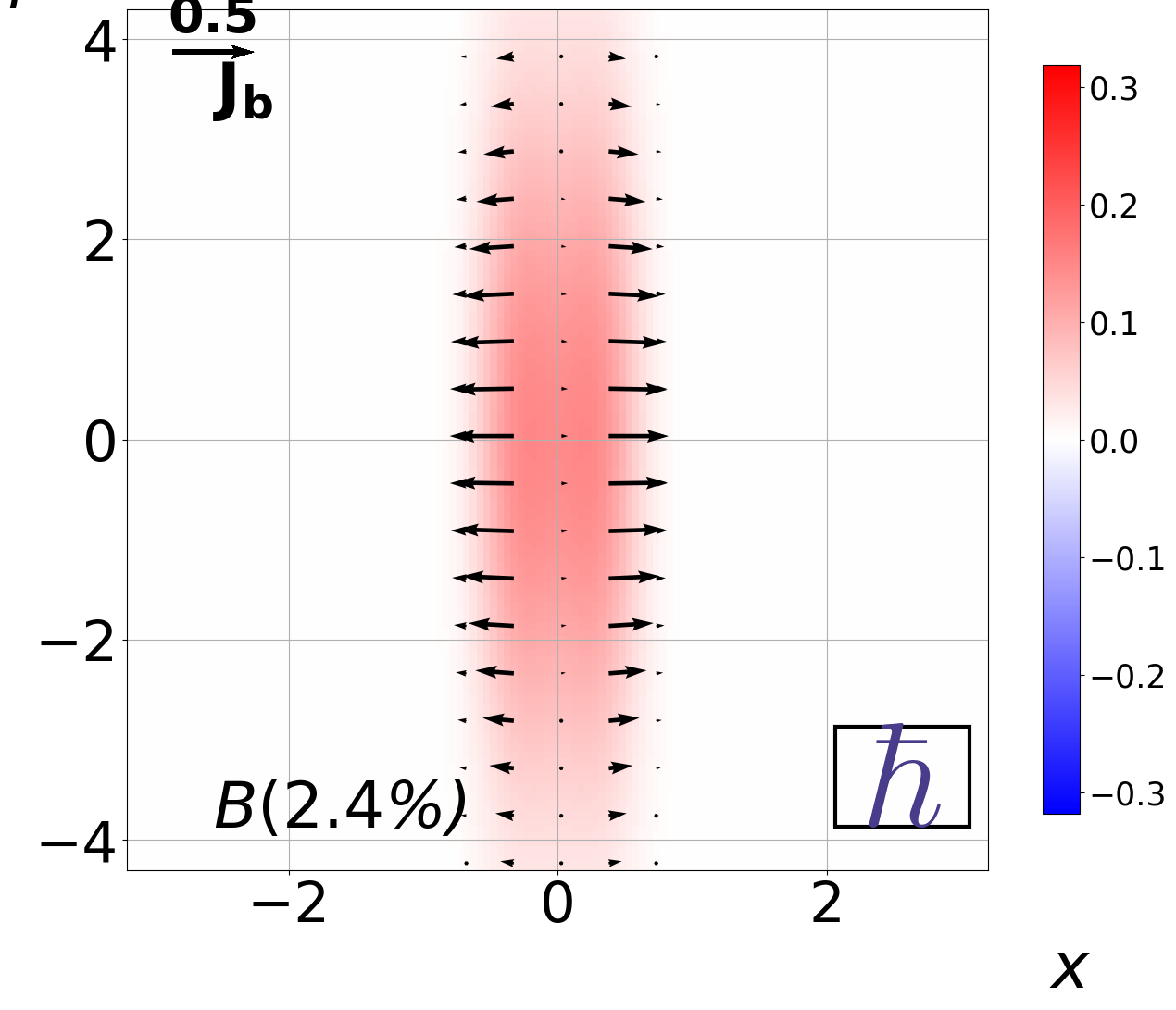}
  \\
  \includegraphics[width=3.8cm,height=3.2cm]{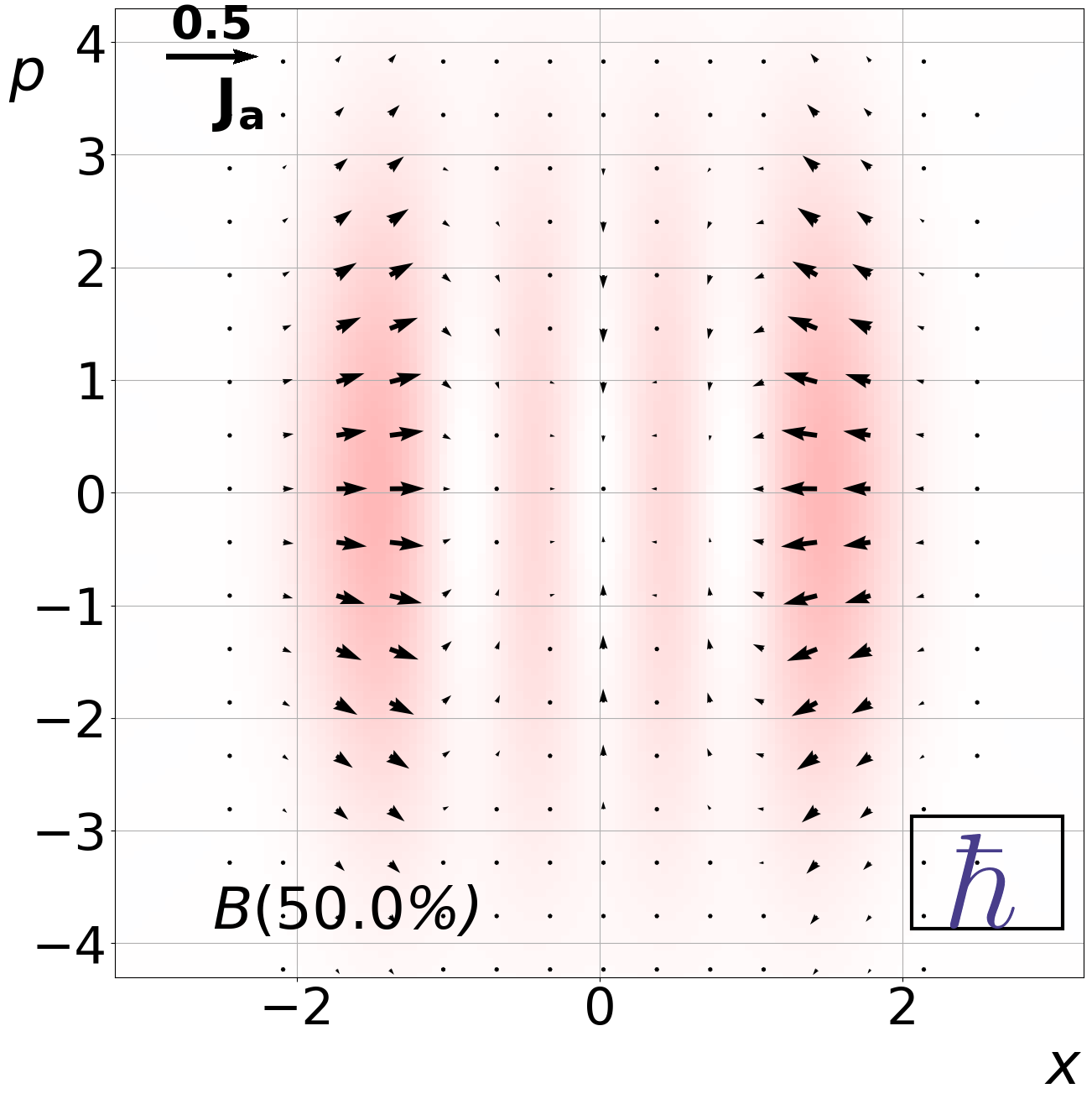}
  \includegraphics[width=4.5cm,height=3.4cm]{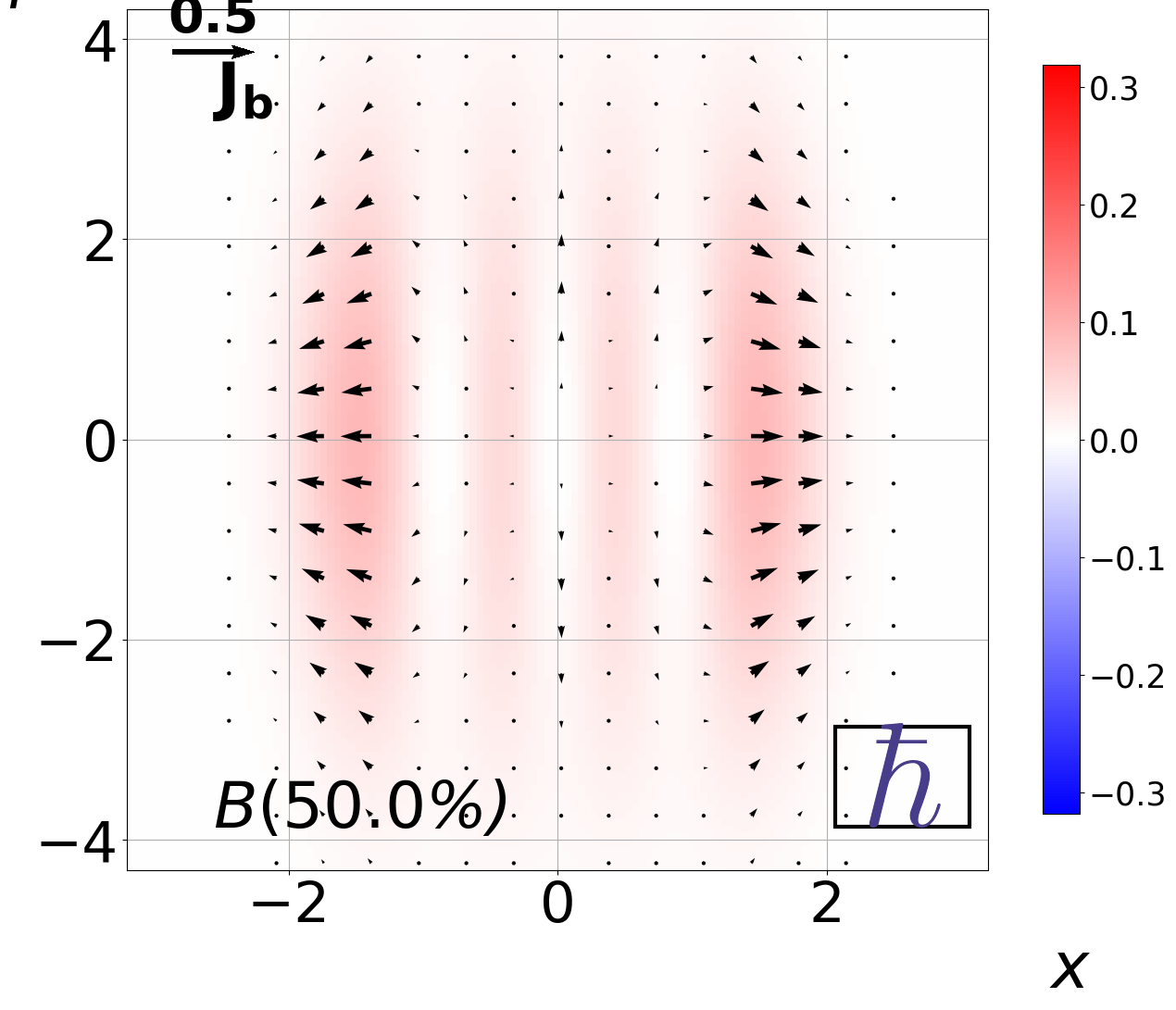}
  \caption{Layout as in Fig.~\ref{fig:SqeFock1}.  Initial state a three-photon Fock-state
    and squeezed vacuum:
    $|\psi_a \rangle {} |\psi_b \rangle = |3\rangle\otimes |z=1.2, \theta = 0\rangle$.
    \label{fig:Sq_fock3}}
\end{figure}

\begin{figure}[t] \centering
  \includegraphics[width=3.8cm,height=3.2cm]{Figures/sq0_n3_A.pdf}
  \includegraphics[width=4.5cm,height=3.4cm]{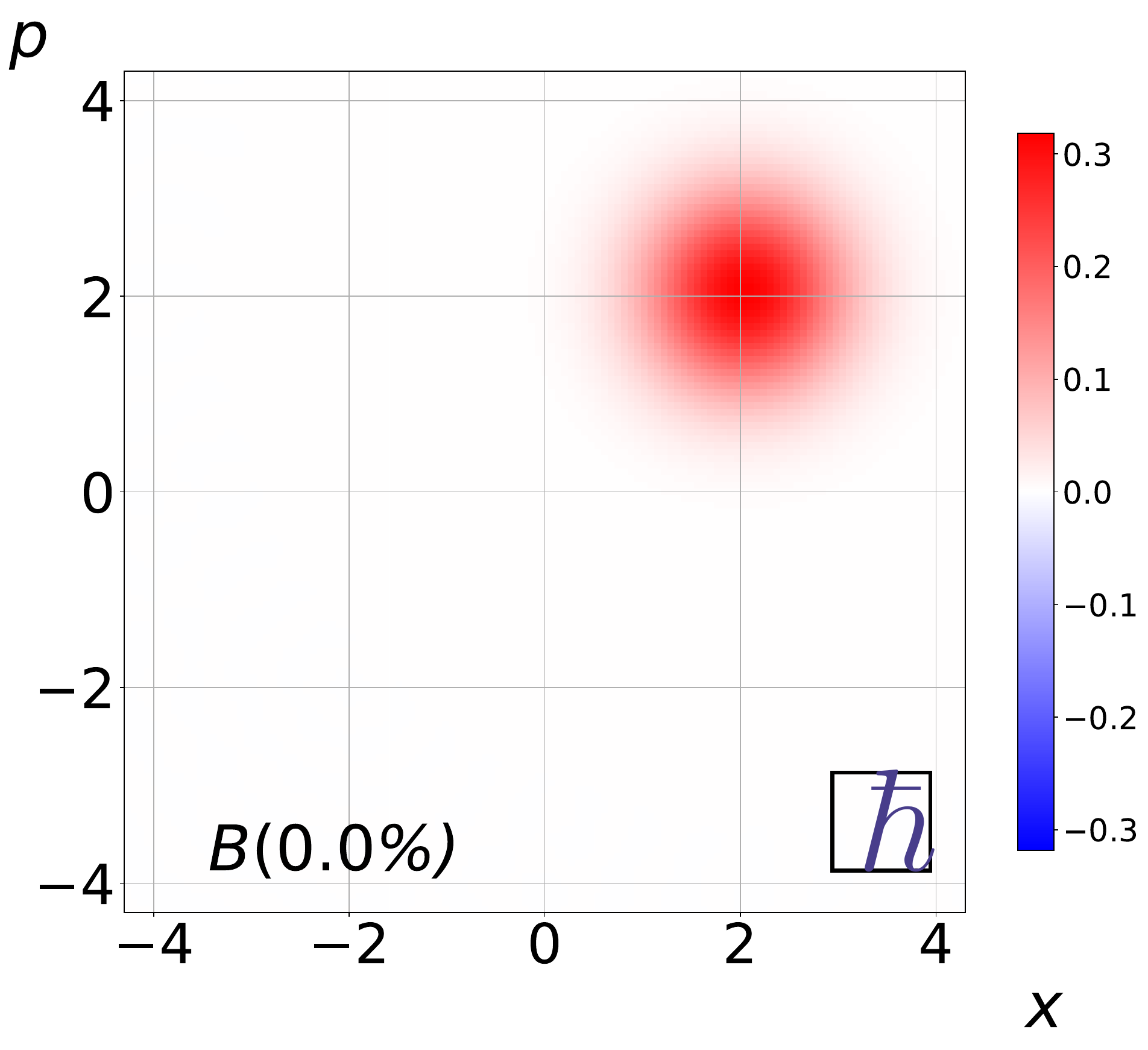}
  \\
  \includegraphics[width=3.8cm,height=3.2cm]{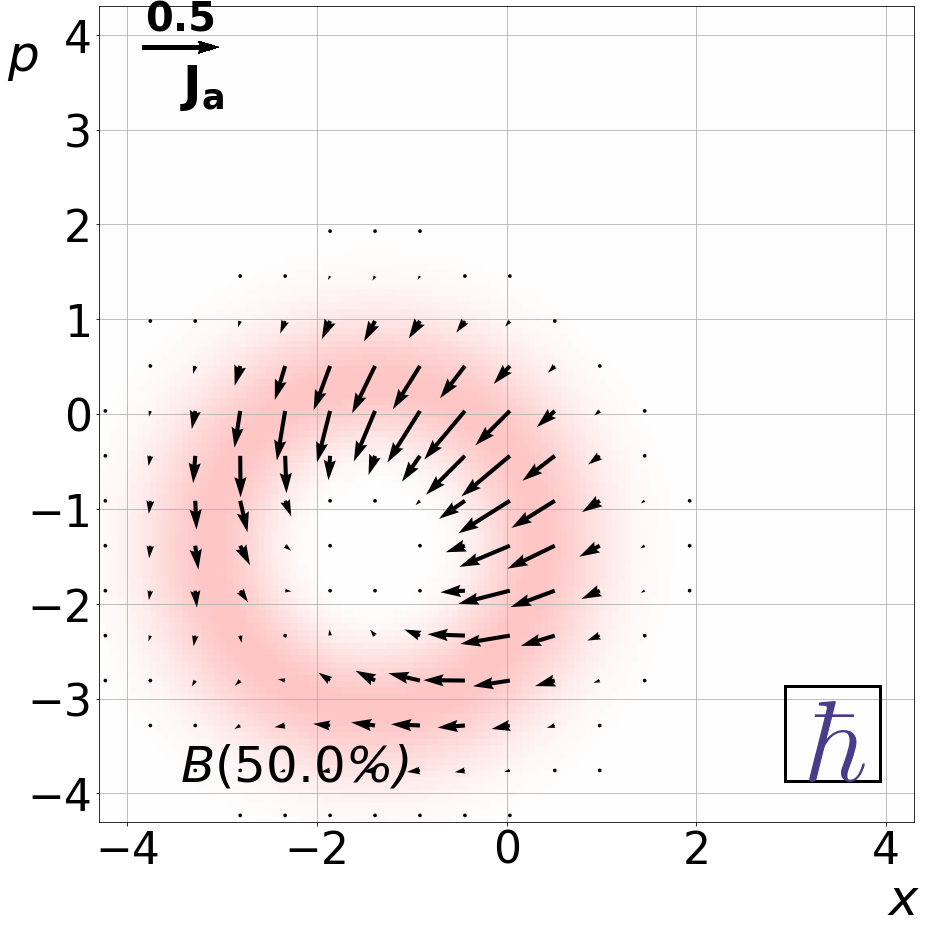}
  \includegraphics[width=4.5cm,height=3.4cm]{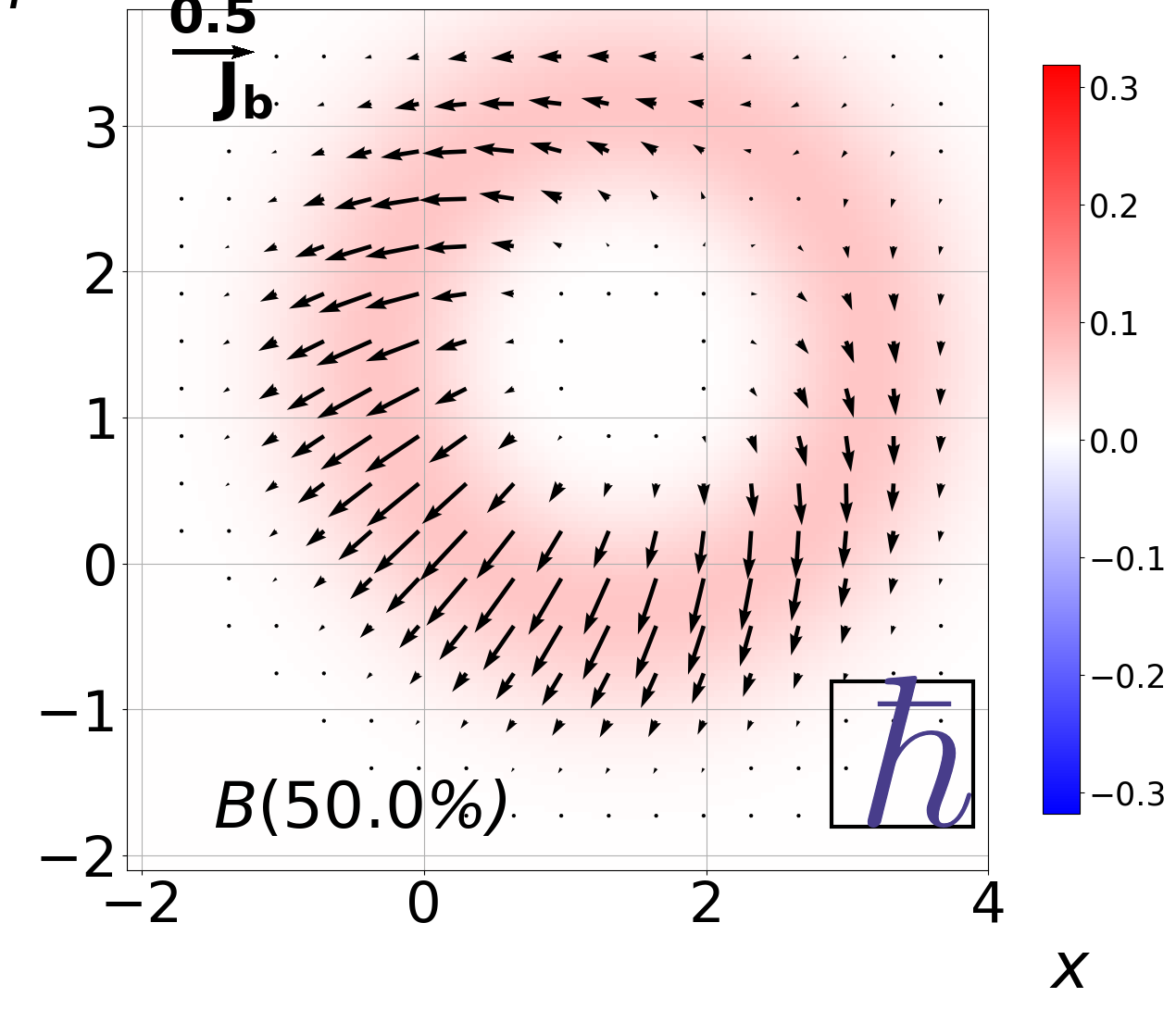}
  \\
  \includegraphics[width=3.8cm,height=3.2cm]{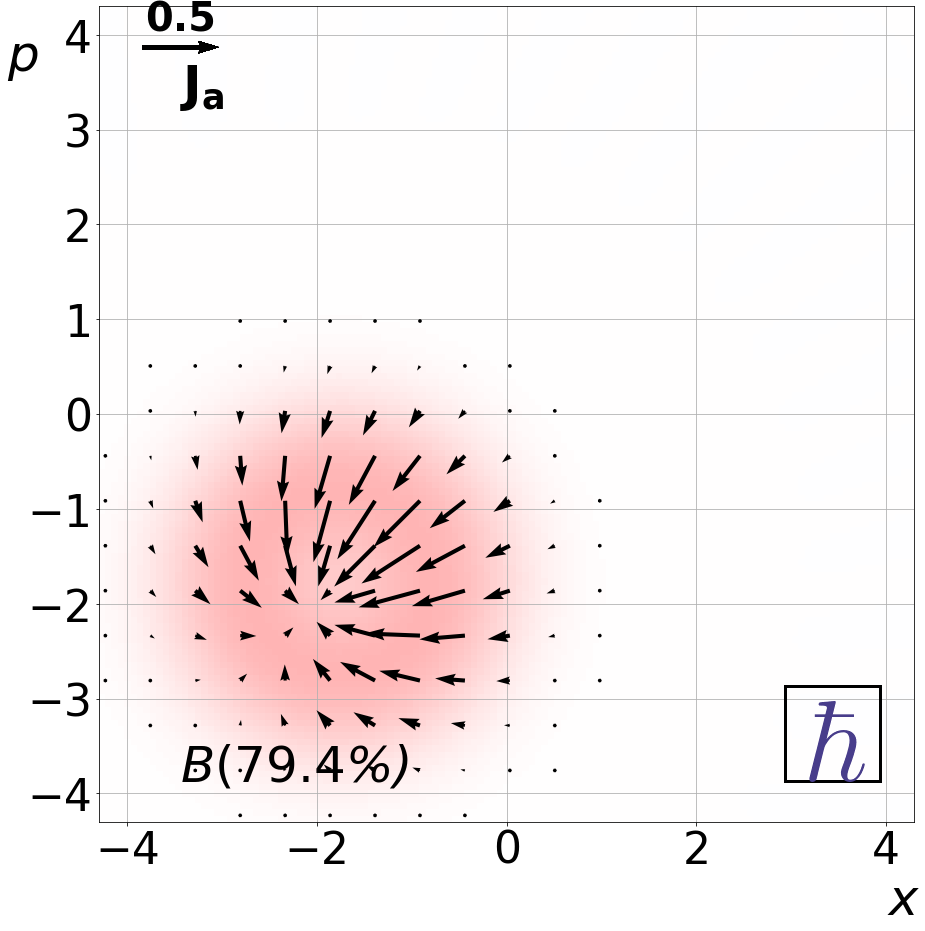}
  \includegraphics[width=4.5cm,height=3.4cm]{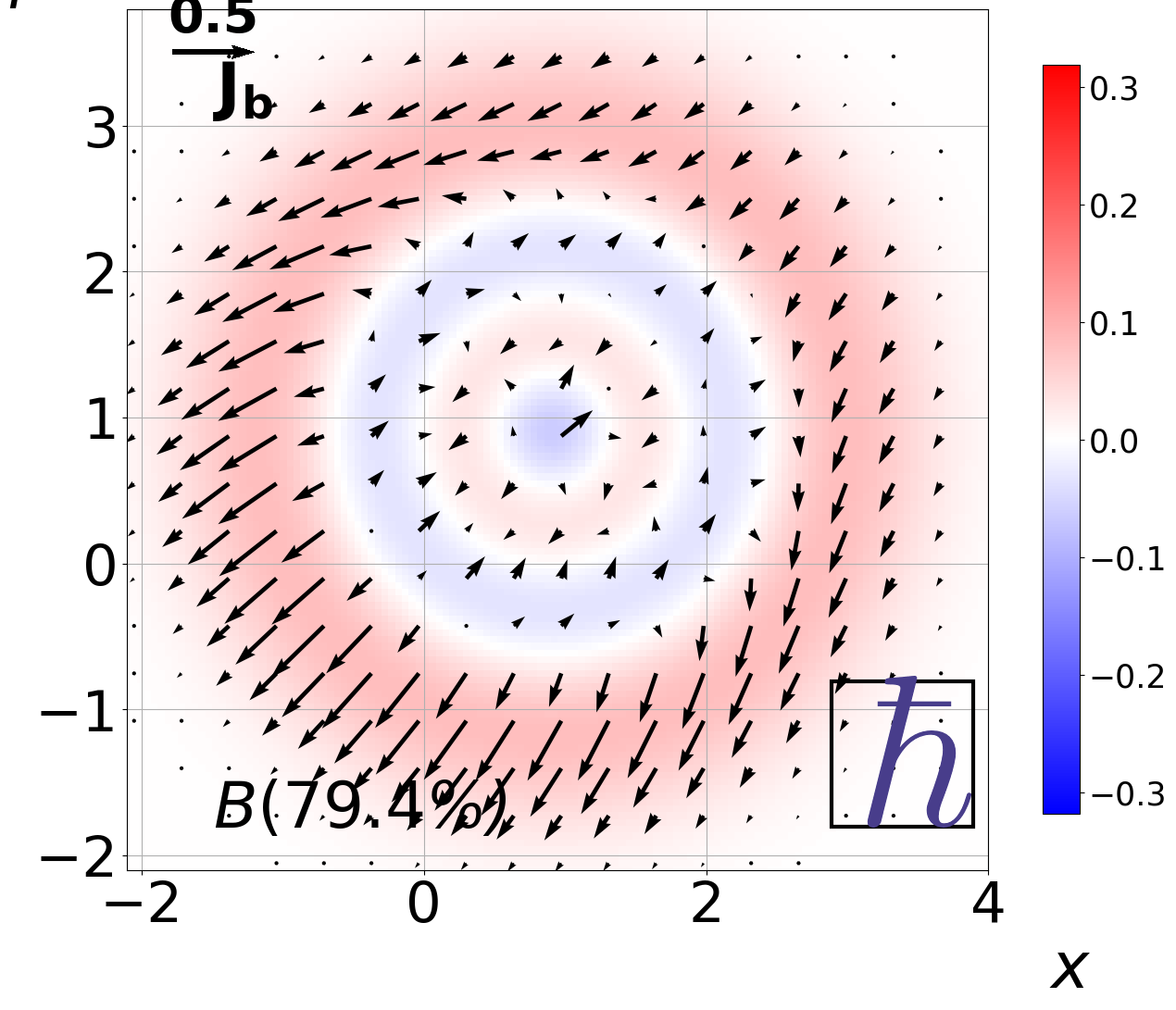}
  \caption{Layout as in Fig.~\ref{fig:SqeFock1}.  Initial state a three-photon Fock-state and coherent state:
    $|\psi_a \rangle {} |\psi_b \rangle = |3\rangle\otimes |\alpha = 2(1+{\rm i})/\sqrt{2}\rangle$.
    \label{fig:Coh_fock3}}
\end{figure}

\subsection{Energy and Purity Changes\label{sec:EnergyPurity}}

Fig.~\ref{fig:SqeFock1} illustrates that the modes can exchange energy, mode~$a$ loses and~$b$ gains
energy. The $\VEC{J}$-fields are integrated yielding white-yellow-blue field lines which expand and
converge, hinting at the fact that \ps volumes are not conserved~\cite{Oliva_PhysA17}.

Let us consider roughly balanced beam splitters for mixing modes, with initial states such as two
single photons or different squeezed states.  It is known that for such strongly mixing beam
splitters the modes become strongly entangled with each other (Fig.~\ref{fig:HOM} describes
Hong-Ou-Mandel interference~\cite{HongOuMandel87}), whilst the individual modes become fairly impure
by themselves: throughout \ps the distributions $W_{a \setminus b}$ become positive and widely
spread-out, see Figs.~\ref{fig:HOM} and~\ref{fig:SquSqu}.

Here, the Wigner current patterns can reveal further details of such purity destroying
dynamics~\cite{Kim__PRA02}. In the top row of Fig.~\ref{fig:HOM} the central inflow fills the
negative region around the origin without affecting the positive torus very much, leading to fully
mixed states (middle row).

\subsection{Changes in \ps Volumes\label{sec:VolumeChanges}}

It is known that the formation of `negative regions', where the Wigner distribution has
negative values, implies that \ps volumes are not conserved. Here, this is underpinned in a visual
way by the observation of radially out- or inward pointing star-current patterns, see
Figs.~\ref{fig:SqeFock1} and~\ref{fig:HOM}. According to Gauss' law (or flux theorem) this implies
the existence of sources or sinks of the local probability current~$\VEC J$. Such local changes of
Wigner's quasi-probability density distribution are alien to classical physics but required in the
quantum case, also for conservative systems~\cite{Oliva_PhysA17}.

In the case of Fig.~\ref{fig:HOM} we can explicitly consider this lack of volume conservation. The
current fields are strictly radial but the insets in the panels of the Right Column of
Fig.~\ref{fig:HOM} show that the magnitude of $\VEC w$ does \emph{not} drop off with the inverse
radius $R^{-1}=1/\sqrt{x^2+p^2}$. Thus, $\VEC{\nabla} \cdot \VEC{w} \neq 0$, quantum \ps volumes are
not conserved. This is not necessarily surprising since we look at a subsystem of a coupled system
and tracing out the partner system means the subsystem is not governed by a conservative
hamiltonian.  But what is impossible in classical physics is the occurrence of singular velocity
fields~\cite{Oliva_PhysA17}, see explanation in Sect.~\ref{sec:ClassicalContrast} and insets in
Right Column of Fig.~\ref{fig:HOM}.

In the next subsection we explain the occurrence of singular \ps volume changes and their connection
with mode entanglement.

\subsection{Negative Values of $W$ from entanglement\label{sec:Entanglement_J_W}}

For beam splitters, the Liouville condition for \ps volume conservation can be \emph{maximally
  viola\-ted}, \ps volume changes can be singular: $|\VEC{\nabla} \cdot \VEC{w}| = \infty$.  We
explained the possibility for such singularities occurring in terms of the lifting of the degeneracy
in the formation of zeros of $\VEC J$ and $W$, see subsection~\ref{sec:ClassicalContrast}.

Yet, at a first careless glance, it is surprising to find singularities in $\VEC{w}$ for
hamiltonians~(\ref{eq:H_M}) \emph{bilinear} in $x$ and $p$, since only the `classical' looking
linear term~\cite{Takabayasi_PTP54,Oliva_PhysA17} of Moyal-bracket~(\ref{EqMoyalBraket}) is
present in evolution~equation~(\ref{eq:Moyal_M_TrA_}).

Why do we find singularities in $\VEC{w}$ (such as those shown in the insets in the Right Column of
Fig.~\ref{fig:HOM})?

The reason is mode entanglement~\cite{Kim__PRA02,Wang__PRA02}:
\\
The system can form zeros in $W$ while, simultaneously, $\VEC J$ does not vanish.  Again, the
degeneracy in the formation of zeros of $\VEC J$ and $W$ are lifted, causing singularities
in~$\VEC w$. As a generic example, let us consider the $x$-component of
${\VEC J}_b$~(\ref{eq:_J_B_explicit}). It is proportional
to~${\rm Tr}_a \{ x_{a} W_{ab} \} (x_b, p_b, \tau)$ and this expression is generally \emph{not
  proportional} to $W_b$, if modes $a$ and $b$ are \emph{entangled}. Hence, when $W_b$ forms a
negative region it vanishes on the boundary of that region and there $\VEC{w}_b = \VEC{J}_b / {W}_b$
can form singularities.

Such singular \ps volume changes, or rather, their signature --nonzero ${\VEC J}$, whilst the density
$W$ vani\-shes-- have so far not been observed experimentally.

\subsection{Phases and Fluctuations\label{sec:PhasesFluctuations}}

We remark that it might seem surprising to note that Eqns.~(\ref{eq:_J_A_explicit})
and~(\ref{eq:_J_B_explicit}) for ${\VEC J}_a$ and ${\VEC J}_b$ are of the same mathema\-tical form if
the indeces $a$ and $b$ are swapped, except for the presence of a minus sign for ${\VEC J}_a$.  Yet,
${\VEC J}_a$'s and ${\VEC J}_b$'s current plots in Fig.~\ref{fig:HOM} are identical. This is due to
the fact that the definitions of the beam splitter modes carry rela\-tive phases which are imprinted
within $W_{ab}$, correctly compensating for ${\VEC J}_a$'s minus sign.

Fig.~\ref{fig:SquSqu} shows that the squeezed states affect each other in a geometrically
transparent fashion: the large fluctuations in the direction of the anti-squeezed quadrature
dominate the direction in which the current broadens the state in the other mode.  This type of
geometrical reasoning also explains qualitatively how the symmetry breaking seen in the left column of
Fig.~\ref{fig:Sq_fock3} arises: the non-classical \ps-interference fringes~\cite{Schleich_01} in the
$a$-mode's Fock state wash out more quickly in the $p$-direction than the $x$-direction since the
$b$-mode is squeezed in its $x$-direction and anti-squeezed in~$p$.

We remark that, according to Eqns.~(\ref{eq:_J_B_explicit}) and~(\ref{eq:_J_A_explicit}), the beam
splitter~(\ref{eq:U_M}) only mixes the positions $x_a$ and $x_b$ with each other, and the momenta
$p_a$ and $p_b$ with each other. There is no cross mixing of positions with
momenta~\cite{Leonhardt_PRA93}.

It is not always obvious how Eqns.~(\ref{eq:_J_B_explicit}) and~(\ref{eq:_J_A_explicit}) for
${\VEC J}_b$ and ${\VEC J}_a$ give rise to the described phenomena. After all, the relative phases
that govern the direction of the dynamics (another example, for a tunnelling scenario, can be found in
the discussion around Fig.~2 of Ref.~\cite{Ole_PRL13}) are `hidden' in the expressions for $W_{ab}$
of Eqns.~(\ref{eq:_J_B_explicit}) and~(\ref{eq:_J_A_explicit}).  In the case of
Fig.~\ref{fig:SquSqu}, for example, initially (for small beam splitter reflectivities) the relative
phases are such that the distributions become smeared out but later on (for much greater
reflectivities) the process, instead, narrows the distributions. All this, whilst the form of
Eqns.~(\ref{eq:_J_B_explicit}) and~(\ref{eq:_J_A_explicit}) remain unchanged, but the expressions
for $W_{ab}$ have changed.

The Wigner current inverts its direction when~$W$ is negative~\cite{Oliva_Kerr_18}, such inversions
(current moves against the prevailing direction) are clearly displayed in the lower right panels of
Figs.~\ref{fig:SqeFock1} and~\ref{fig:Coh_fock3}.

\section{Conclusions\label{sec:Conclusion}}

We have shown how the Wigner current for coupled systems can be determined after tracing out one or
the other subsytem. We specified this for the case of variable beam splitters and highlighted
various aspects of the corres\-ponding \ps dynamics. It is straightforward to apply the calculations
in Sec.~\ref{sec:WignerJ2mode} to other forms of interaction hamiltonians coupling two systems.

We emphasized that the effective Moyal brackets descri\-bing the conditional beam splitter dynamics
show features familiar from systems coupled to environments far from thermal equilibrium: energy or
purity are not conserved in each subsystem. Energy and purity can go up or down.

We specifically highlighted aspects of the quantum \ps dynamic which our analysis has unearthed as
being experimentally accessible when using variable beam splitters and which, in quantum optics, can
be difficult to generate and experimentally investigate other\-wise. These are: the formation of
negative \mbox{areas} of the Wigner distributions, the inversion of Wigner's \ps currents associated
with these negative \mbox{areas}, and the pronounced violation of \ps volume
conservation~\cite{Oliva_PhysA17} in quantum dynamics, which are all tied to the entanglement
between the two mode~\cite{Kim__PRA02,Wang__PRA02}.

Traditionally, states and how they change over time is studied, here we suggest to additionally use
the vector fields ${\VEC J}_a$ and ${\VEC J}_b$. It allows us to apply \emph{geometrical reasoning}
to extract quantitative and qualitative information of how their interaction drives the system
dynamics. This is one of the main findings of this work and one of its main motivations.

We hope that the approach laid out in this work can also lead to new ideas about how to tailor
hamiltonians~\cite{Braasch_PRA19} and states (here, depending on the `other mode' that is traced
over) to generate interesting new quantum states and quantum dynamics.  We anticipate that the
geometrical reasoning demonstrated here can also be applied in other multimode cases with other
couplings and will help with the detailed analysis of the dynamics of such systems, based on the
analysis of Wigner's \ps current~$\VEC J$.

\section*{Acknowledgements}
This work is partially supported by the Ministry of Science and Technology of Taiwan (Nos
112-2123-M-007-001, 112-2119-M-008-007, 112-2119-M-007-006), Office of Naval Research Global, the
International Technology Center Indo-Pacific (ITC IPAC) and Army Research Office, under Contract
No. FA5209-21-P-0158, and the collaborative research program of the Institute for Cosmic Ray
Research (ICRR) at the University of Tokyo.

\section*{Disclosures} The authors declare no conflicts of interest.


\begin{thebibliography}{50}
\bibitem [{for()}]{formulationsQuantumTheoryfootnote}%
  \BibitemOpen
  \href@noop {} {}\bibinfo {note} {There are several different formulations of
  quantum theory, Schr{\"o}dinger's, Heisenberg's, the \ps formulations due to
  Wigner, Moyal and Groenewold (and their offshoots due to Glauber and
  Sudarshan, Husimi and others), Feynman's path integrals, the de~Broglie--Bohm
  and many-worlds interpretation, spontaneous collapse models and others. Here
  we only use Wigner's formulation~\cite{WignerSpecialfootnote}.}\BibitemShut
  {Stop}%
\bibitem [{Wig()}]{WignerSpecialfootnote}%
  \BibitemOpen
  \href@noop {} {}\bibinfo {note} {Out of all quantum phase space
  distributions~\cite{Cahill_PR69b} only Wigner's~\protect{\cite{Hillery_PR84}}
  yields the correct projections in Schr{\"o}dinger's position
  $\varrho(x,x,t)=\int dp \; W(x,p,t)$~and momentum densities $\tilde
  \varrho(p,p,t)=\int dx \; W(x,p,t)$,~whilst giving the overlap between states
  in the simple~form $|\langle \psi_1(x,t) | \psi_2(x,t) \rangle |^2 = 2 \pi
  \hbar \iint dx dp $$\, W_1(x,p,t) $$ W_2(x,p,t)$. We therefore consider
  Wigner's approach~\cite{Wigner_PR32} special and intuitive, we use it
  exclusively}\BibitemShut {NoStop}%
\end{thebibliography}
%


\end{document}